\documentclass[11pt, twoside]{article}
\pdfoutput=1
\usepackage{a4wide,amssymb,cite,graphicx}
\usepackage{epsfig}
\usepackage[usenames,dvipsnames]{color}
\usepackage{slashed}

\newcommand{\be}{\begin{equation}}
\newcommand{\ee}{\end{equation}}
\newcommand{\bea}{\begin{eqnarray}}
\newcommand{\eea}{\end{eqnarray}}

\begin{document}

\color{black}

\begin{flushright}
CERN-PH-TH/2012-340 \\
KIAS-P12081
\end{flushright}

\vspace{0.5cm}
\begin{center}
{\huge\bf\color{black}  Interplay between Fermi gamma-ray  \\[3mm]  lines  and collider searches}\\
\bigskip\color{black}\vspace{1.0cm}{
{\large\bf Hyun Min Lee$^{1}$, Myeonghun Park$^{2}$ and Ver\'onica Sanz$^{2,3}$}
\vspace{0.5cm}
} \\[8mm]

{\it $^1$School of Physics, KIAS, Seoul 130-722, Korea.  }\\
{\it $^2$Theory Division, Physics Department, CERN,  CH--1211 Geneva 23,  Switzerland.}\\
{\it $^3$Department of Physics Astronomy, York University, Toronto, ON M3J 1P3, Canada. } \\
\end{center}
\bigskip
\centerline{\large\bf Abstract}
\begin{quote}\large
We explore the interplay between lines in the gamma-ray spectrum and LHC searches involving missing energy and photons. As an example,
 we consider a singlet Dirac fermion dark matter with the mediator for Fermi gamma-ray line at 130 GeV. A new chiral or local $U(1)$ symmetry makes weak-scale dark matter natural and provides the axion or $Z'$ gauge boson as the mediator connecting between dark matter and electroweak gauge bosons. In these models, the mediator particle can be produced in association with a monophoton at colliders and it produces large missing energy through the decays into a DM pair or $ZZ, Z\gamma$ with at least one $Z$ decaying into a neutrino pair.  We adopt the monophoton searches with large missing energy at the LHC and impose the bounds on the coupling and mass of the mediator field in the models. We show that the parameter space of the $Z'$ mediation model is already strongly constrained by the LHC 8 TeV data, whereas a certain region of the parameter space away from the resonance in axion-like mediator models are bounded. We foresee the monophoton bounds on the $Z'$ and axion mediation models at the LHC 14 TeV.

\end{quote} 

\thispagestyle{empty}

\normalsize

\newpage

\setcounter{page}{1}

\section{Introduction}

%dark matter

%Fermi gamma-ray line
Recently, from the Fermi LAT data \cite{fermi,dwarfgalaxy}, there has been an interesting observation of the gamma-ray line at $130\,{\rm GeV}$ \cite{weniger,otheranalysis}, which may call for an explanation from physics beyond the Standard Model (SM) \footnote{The preliminary result of the Fermi LAT collaboration has confirmed the $130\,{\rm GeV}$ gamma-ray line, which is shifted to $135\,{\rm GeV}$ with the reprocessed four-year data though \cite{FermiSymposium}.  There is a concern with a bump observed in the Earth limb data at the similar energy \cite{earthlimb,FermiSymposium}. }. Although dark matter(DM) has not been found directly by underground experiments such as XENON \cite{xenon}, it is an interesting possibility that  the Fermi gamma-ray line is produced by the DM annihilation into monochromatic photon(s), such as a photon pair or $Z\gamma$ or $h\gamma$. 
Then, the DM annihilation cross section for a pair of monochromatic photons is required to be around $4-8\%$ of the thermal cross section $\langle\sigma v\rangle=3\times 10^{-26}{\rm cm}^3/{\rm s}$ \cite{weniger}. But, the DM annihilation into monochromatic photon(s) must be loop-suppressed, because dark matter is charge-neutral. Thus, it is challenging to find a model for suppressing the tree-level annihilation channels into the SM particles while the loop-induced annihilation channels are sizable \cite{othermodels,scalarDM,axion1,axion2,zprimeCS,zprimeTquark}.
For instance, one needs a large  coupling between dark matter and a new charged particle running in loops or rely on a resonance effect in the presence of a mediator.

%Models with mediator
We consider a singlet Dirac fermion dark matter whose mass can be of weak scale naturally due to a new chiral symmetry or gauge symmetry.
If there is no direct interaction between DM and new charged particles, we need a mediator field connecting dark matter to the SM particles through the s-channel. Since the DM annihilation must be s-wave to be a relevant process at present, the mediator is either a pseudo-scalar (axion-mediation) \cite{axion1,axion2} or a vector ($Z'$ mediation) \cite{zprimeCS,zprimeTquark}.  
In this paper, we revisit two representative models for a singlet Dirac fermion dark matter with the axion scalar or $Z'$ gauge boson mediators, from the perspective of the collider physics.  First, we review the mechanisms to generate the Fermi gamma-ray line in the considered models: DM annihilation into a pair of monochromatic photons in axion mediation and DM annihilation into a single monochromatic photon due to $Z\gamma$ or $h\gamma$ channel in $Z'$ mediation. In axion mediation, a weak-scale axion couples to dark matter and also generates $U(1)$ Peccei-Quinn(PQ) anomalies depending on the SM representation and PQ charge of a heavy fermion running in triangle loops. In this case, the axion resonance enhances the DM annihilation cross section into a photon pair. In $Z'$ mediation, dark matter is charged under an extra local $U(1)$ with the Chern-Simons(CS) interactions to the electroweak gauge bosons. Thus, depending on the value of the CS coupling, the DM annihilation cross section into a single photon can be enhanced with or without a resonance. We recap the consequences of the two models in view of the Fermi gamma-ray line, including the relic density condition. In particular, we first include a new DM annihilation channel into $WW$ in $Z'$ mediation in the presence of an extra CS term for the W-boson. This new channel provides a freedom to take the $Z'$-boson mass even above twice the DM mass, being consistent with the relic density.

%monophoton search and bounds on models
Searches for a monophoton or monojet with large missing energy at the Large Hadron Collider (LHC) have become a powerful constraint on dark matter and large extra dimensions \cite{CMS-monoa,ATLAS-monoa}. In the effective operator approach for dark matter interactions with the SM quarks, a DM pair can be produced directly from the scattering of an initial quark pair so a monophoton radiated by the initial quark is recoiled against the missing energy \cite{monogsearches}. In the DM mediator models that we consider,  however,  the effective coupling between dark matter and the SM quarks is absent after the mediator is integrated out  \footnote{One can introduce the scalar partner of the axion mediator with Higgs portal coupling in axion mediation \cite{axion1} or the kinetic mixing between $Z'$ and $Z$ bosons in $Z'$ mediation \cite{zprimeTquark}, for which the effective DM couplings to the SM quarks appear at tree level. Then, those couplings can be constrained by DM direct detection experiments.}. Thus, dark matter cannot be produced directly from an initial quark pair, but instead the mediator field is produced in association with a monophoton. If the mediator decays into a DM pair, it is possible to impose the bound of the monophoton search with large missing energy at the LHC \footnote{In the case of a scalar dark matter, the DM annihilation into a photon pair is induced by loops containing a charged scalar \cite{scalarDM}. In this case, it would be possible to have a direct production of a monophoton in association with a scalar DM pair too.}. Below the mass threshold of a DM pair, we rely on the mediator decay modes into $ZZ$ or $Z\gamma$ and we expect the high efficiency of the photon cut and missing energy if at least one $Z\rightarrow \nu {\bar\nu}$ is required.
We consider the collider bounds on the coupling and mass of the mediator field from the monophoton searches at the LHC 8 TeV and discuss the prospects at the 14 TeV upgrade. 
In axion mediation, the axion production cross section is suppressed due to a small coupling between the axion and the electroweak gauge bosons, but a certain region of the parameter space, away from the resonance, can be constrained by the LHC 8 TeV.
On the other hand, in $Z'$ mediation, if the CS coupling between the $Z'$-boson and the electroweak gauge bosons is sizable, the $Z'$ production cross section is large enough to be accessible in the monophoton search already at the LHC 8 TeV.

The paper is organized as follows.
We begin with the description of the dark matter models and discuss the constraints on the model parameters from Fermi gamma-ray line and DM relic density. Then we present the production cross sections of the mediator fields at the LHC in both DM models and set the collider bounds on the models from the monophoton searches at the LHC 8 TeV and 14 TeV.  Finally, conclusions are drawn. There are two appendices showing the detailed computations of the DM annihilation cross sections and decay rates, and the mediator production cross sections.

\section{Models for Fermi gamma-ray line}

We consider fermion dark matter models with axion scalar or $Z'$ gauge boson mediators
and discuss the constraints on the models coming from Fermi gamma-ray line and relic density. 
The stability of fermion dark matter is ensured by the $Z_2$ symmetry, $\chi\rightarrow -\chi$, which can be the remnant of the breaking of PQ symmetry or local $U(1)$.

\subsection{Axion-mediated fermion dark matter}

First, in axion mediation \cite{axion1,axion2}, after the $U(1)$ PQ symmetry is spontaneously broken, 
the effective action for a Dirac fermion dark matter $\chi$ and a pseudo-scalar (axion) $a$ can be written in the following,
\bea
{\cal L}_{\rm axion}&=& i{\bar\chi}\gamma^\mu\partial_\mu\chi- m_\chi {\bar\chi}\chi +\frac{1}{2}(\partial_\mu a)^2-\frac{1}{2}m^2_a a^2-\frac{1}{4}F^i_{\mu\nu} F^{i\mu\nu} \nonumber \\
&&+\frac{1}{\sqrt{2}}\lambda_\chi i a \, {\bar\chi} \gamma^5 \chi + \sum_{i=1,2}\frac{c_i\alpha_i}{ 8\pi  v_s}\, a F^i_{\mu\nu} {\tilde F}^{i\mu\nu}
\eea
where $c_1, c_2$ are constant parameters given by $c_1={\rm Tr}( q_{PQ}Y^2)$ and $c_2={\rm Tr}(q_{\rm PQ}l(r))$ with $q_{\rm PQ} (Y)$ being PQ charge (hypercharge) and $l(r)$ being the Dynkin index of representation $r$ of the extra lepton under the $SU(2)_L$. When dark matter gets mass by the VEV of the scalar partner of the axion, the dark matter coupling is constrained by $\lambda_{\chi}=\sqrt{2}\, m_\chi/v_s$. 
The effective axion interactions can be rewritten in terms of physical electroweak gauge bosons,
\bea
{\cal L}_{\rm anomaly}&=&c_{\gamma\gamma}a\,\epsilon_{\mu\nu\rho\sigma} F^{\mu\nu}_{\gamma}F^{\rho\sigma}_{\gamma}+ c_{Z\gamma}a\,\epsilon_{\mu\nu\rho\sigma} F^{\mu\nu}_{Z}F^{\rho\sigma}_{\gamma} \nonumber \\ 
&&+c_{ZZ}a\,\epsilon_{\mu\nu\rho\sigma} F^{\mu\nu}_{Z}F^{\rho\sigma}_{Z}+c_{WW}a\,\epsilon_{\mu\nu\rho\sigma} F^{\mu\nu}_{W}F^{\rho\sigma}_{W}\label{anomalyint}
\eea
where
\bea
c_{\gamma\gamma}&=& \frac{1}{16\pi  v_s}(c_1\alpha_1 \cos^2\theta_W+c_2\alpha_2 \sin^2\theta_W), \\
c_{Z\gamma}&=& \frac{1}{16\pi v_s} (c_2\alpha_2 -c_1\alpha_1)\sin(2\theta_W), \\
c_{ZZ}&=& \frac{1}{16\pi v_s}(c_2\alpha_2  \cos^2\theta_W+c_1 \alpha_1 \sin^2\theta_W),\\
c_{WW}&=& \frac{ c_2\alpha_2}{8\pi v_s}.
\eea

We note that the axion mass appears to break the PQ symmetry explicitly, but it can come from a PQ invariant higher dimensional operator, such as $\frac{1}{M^2_P} S^2 \Phi^4$, where a complex scalar $S=s+ia$ contains our weak-scale axion as the imaginary part and $\Phi$ is another complex scalar breaking the PQ symmetry spontaneously at high scale by $\langle\Phi\rangle=v_\Phi$ with $10^9\,{\rm GeV}\lesssim v_\Phi \lesssim 10^{12}\,{\rm GeV}$ and consequently containing the invisible axion. In the presence of the higher dimensional operator, there is a mixing between the invisible axion $a_\Phi$ and our weak-scale axion $a_s$. For the $S$ singlet with PQ charge $-2$, the higher dimensional operator leads to $\frac{a_s}{v_s}\simeq a_2/v_s-2a_1/v_\Phi$ and $\frac{a_\Phi}{v_\Phi}\simeq a_1/v_\Phi + 2v_s a_2 / v^2_\Phi$ for $v_\Phi\gg v_s$, in terms of  the mass eigenstates  $a_{1,2}$ of invisible and weak-scale axions. Thus, the couplings of $a_\Phi$ to extra heavy quarks \cite{kimaxion} or the SM quarks \cite{dineaxion}  generate the QCD anomalies dominantly by $a_1$, whereas the couplings of $a_s$ to extra heavy leptons generate the electroweak  anomalies dominantly by $a_2$ \cite{axion2}.

\begin{figure}[t]
\centering%
\includegraphics[width=13cm,natwidth=610,natheight=642]{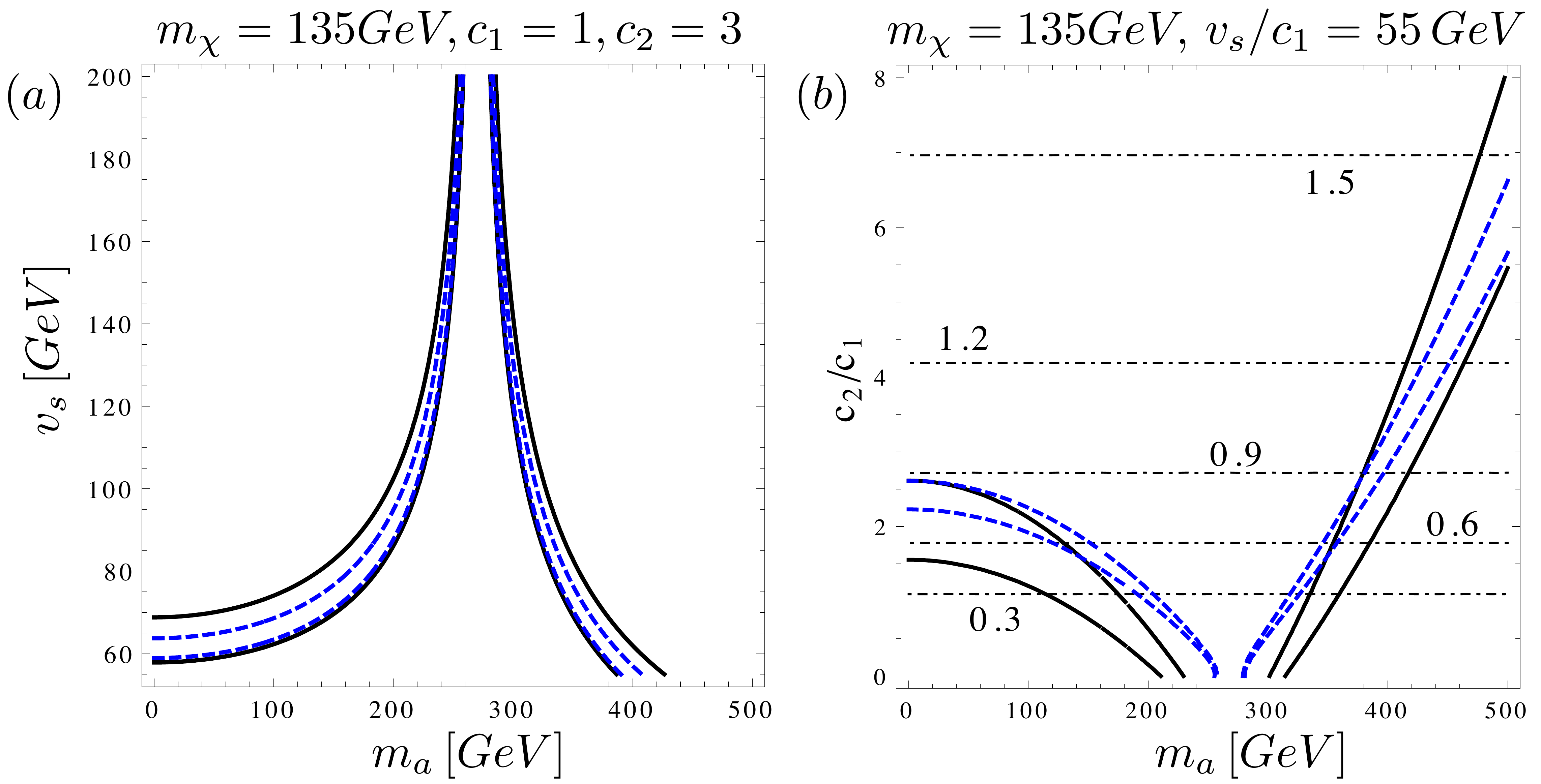} 
\caption{Left: Region between nearby black solid lines is the parameter space of $m_a$ vs $v_s$ satisfying $\langle\sigma v\rangle_{\gamma\gamma}/\langle \sigma v\rangle_0=0.04-0.08$ where $\langle\sigma v\rangle_0=3\times 10^{-26}\,{\rm cm}^3/{\rm s}$ is the thermal cross section. Region between nearby blue dashed lines is the parameter space of $m_a$ vs $v_s$ producing the relic density within WMAP $5\sigma$. Right: The same contour plots for $m_a$ vs $c_2/c_1$. The dot-dashed lines are the contours of  $\langle\sigma v\rangle_{Z\gamma}/(2\langle\sigma v\rangle_{\gamma\gamma})$.
We have set dark matter mass to $m_\chi=135\,{\rm GeV}$ for Fermi gamma-ray line at $135\,{\rm GeV}$.
}\label{fig:axion}
\end{figure}

In this model, the DM annihilation cross sections \footnote{The DM annihilation cross sections for the other channels, $WW$, $ZZ$, $Z\gamma$, are given in Ref.~\cite{axion1}.} into a photon pair and $Z\gamma$ are given by \cite{axion1,axion2}
\bea
\langle\sigma v\rangle_{\gamma\gamma}&=&\frac{ |\lambda_\chi|^2 |c_{\gamma\gamma}|^2}{2\pi }\,  \cdot
\frac{16 m^4_\chi}{(4m^2_\chi-m^2_a)^2+\Gamma^2_a m^2_a},\label{axionxsection} \\
\langle\sigma v\rangle_{Z\gamma}&=&\frac{|\lambda_\chi|^2 |c_{Z \gamma }|^2}{4\pi}\,   \,\cdot
\frac{16 m^4_\chi}{(4m^2_\chi-m^2_a)^2+\Gamma^2_a m^2_a}\Big(1-\frac{m^2_Z}{4 m^2_\chi}\Big)^3
\eea
where the decay rate of the axion is given by
\be
\Gamma_a=\Gamma_a(\gamma\gamma)+\Gamma_a(Z\gamma)+\Gamma_a(ZZ)+\Gamma_a(WW)+\Gamma_a({\bar\chi}\chi)
\ee
with
\bea
\Gamma_a(\gamma\gamma)&=&\frac{m^3_a}{\pi} |c_{\gamma\gamma}|^2, \\
\Gamma_a(Z\gamma)&=& \frac{m^3_a}{2\pi}|c_{Z \gamma }|^2\Big(1-\frac{M^2_Z}{m^2_a}\Big)^3, \\
\Gamma_a(ZZ)&=& \frac{m^3_a}{\pi} |c_{ZZ}|^2\Big(1-\frac{4M^2_Z}{m^2_a}\Big)^{3/2}, \\
\Gamma_a(WW)&=&  \frac{m^3_a}{2\pi}|c_{WW }|^2\Big(1-\frac{4M^2_W}{m^2_a}\Big)^{3/2}, \\
\Gamma_a({\bar\chi}\chi)&=&\frac{|\lambda_\chi|^2}{16\pi} \,m_a\Big(1-\frac{4m^2_\chi}{m^2_a}\Big)^{1/2}.
\label{GammAx}
\eea
The last decay mode into a dark matter pair is allowed only for $m_a> 2m_\chi$.
When the anomaly interactions (\ref{anomalyint}) are generated by an extra charged lepton with PQ charge $q_{PQ}$, electromagnetic charge $Q$, mass $m_f$ and axion coupling $\lambda_f$, the anomaly coefficient for a photon pair is given by $c_1+c_2=q_{PQ} Q^2$ in the decoupling limit of the extra charged lepton.
We can implement the mass dependence of the extra charged lepton with $q_{PQ}=1$ and $Q=-1$ in eq.~(\ref{axionxsection}) by replacing $(c_1+c_2)/v_s$ by $\frac{\lambda_f}{m_f}\,\tau\arcsin^2(1/\sqrt{\tau})$ with $\tau=m^2_f/m^2_\chi$ \cite{axion2}. In the limit with $\tau\gg 1$, we can recover the calculation with effective axion interactions.

In order to explain the observed Fermi gamma ray line by dark matter annihilation,
one needs the cross section of dark matter annihilating into a pair of monochromatic photons to be $\langle\sigma v\rangle_{\gamma\gamma}=(1.27\pm 0.32^{+0.18}_{-0.28})\times 10^{-27}{\rm cm}^3 {\rm s}^{-1}$ for the Einasto profile and
$\langle\sigma v\rangle_{\gamma\gamma}=(2.27\pm 0.57^{+0.32}_{-0.51})\times 10^{-27}{\rm cm}^3 {\rm s}^{-1}$ for the NFW profile, that is, ${\rm Br}({\bar\chi}\chi\rightarrow \gamma\gamma)\simeq 4-8\%$ for thermal dark matter \cite{weniger}.
In axion-mediated dark matter models, the DM annihilation into a photon pair can give rise to Fermi gamma-ray line at $130(135)\,{\rm GeV}$ for $m_\chi=130(135)\,{\rm GeV}$ near the axion resonance with $m_a\sim 2 m_\chi$, for which the loop factor is compensated for a large annihilation cross section. Then, the DM annihilation into  $Z\gamma$ can produce another gamma-ray line at $E_\gamma=114(120)\,{\rm GeV}$.
The branching fraction of the two-photon channel and the decay width of the axion depend on the anomaly coefficients \cite{axion1}. There are microscopic models with heavy vector-like leptons for the PQ anomalies where Fermi gamma-ray line and dark matter relic density can be explained by axion-mediation at the same time \cite{axion2}. 

On the left of Fig.~\ref{fig:axion}, we depict the parameter space compatible with the DM annihilation cross section for $\chi{\bar\chi}\rightarrow \gamma\gamma$ and the relic density condition with WMAP $5\sigma$ bound, on the $m_a$ vs $v_a$ plane, for a set of anomaly coefficients with $c_1=1, c_2=3$, which is the case for a vector-like doublet ($Y=-1/2$) fermion and a triplet ($Y=0$) fermion running in loops \footnote{Extra vector-like leptons can enhance the Higgs-to-diphoton rate by their Yukawa couplings to the Higgs doublet \cite{Hdiphoton,axion2}. The interplay between Fermi gamma-ray line and Higgs diphoton signal in the models with extra vector-like leptons has been discussed recently \cite{axion2}.}. 
On the right of Fig.~\ref{fig:axion}, we show the dependence of the Fermi gamma-ray line and relic density conditions on the ratio of anomaly coefficients for a fixed singlet VEV, $v_s/c_1=55\,{\rm GeV}$. In the same figure, we also overlay the contour lines of the intensity ratio of one-photon to two-photon lines, $r\equiv \langle\sigma v\rangle_{Z\gamma}/(2 \langle\sigma v\rangle_{\gamma\gamma})$.
We note that  when $c_2/c_1\gtrsim 1.5$, there is a parameter space compatible with both Fermi gamma-ray line and relic density, for which the intensity of the one-photon line is comparable to or larger than the one of the two-photon line,  for instance, $r\simeq 1$ for $c_1=1, c_2=3$. 
The presence of two photon lines is an interesting feature of the axion mediated DM models. For the collider study in the later section, we have scanned the parameter space $(v_s,c_1,c_2)$, in the ranges of $55\,{\rm GeV}\leq v_s\leq 200\,{\rm GeV}$ and $0<c_{1,2}\leq10$, for a given axion mass, having in mind the Fermi gamma-ray line and the relic density condition.
We note that the continuum photons coming from $WW, ZZ, Z\gamma$ are consistent with the gamma-ray \cite{dwarfgalaxy} and anti-proton  \cite{antiproton} constraints as well as the shape of the gamma-ray line spectrum \cite{continuum}.

We also note that when the axion is promoted to a complex scalar in a UV completion, the real scalar partner of the axion leads to extra s-channel DM annihilation channels into a pair of the SM particles through Higgs portal term \cite{axion1}. In this case, those extra channels contribute to the thermal cross section at freezeout and determine the relic density correctly together with the axion-mediated channels for general anomaly coefficients, but those channels are p-wave suppressed at present. Then,  thanks to the extra s-channels with the real scalar partner, we are free to choose the anomaly coefficients to be consistent with the Fermi gamma-ray line, without taking into account the relic density condition seriously. However, the mixing between the real scalar and the Higgs boson can be constrained by the XENON100 bound and the Higgs data and Higgs-like boson search at the LHC \cite{axion1}.

\subsection{$Z'$-mediated fermion dark matter}

Second, in $Z'$ mediation, a fermion dark matter has a charge under extra local $U(1)$ and the $Z'$ gauge boson has the effective interaction with a single photon only due to Landau-Yang theorem. The effective action for fermion dark matter $\chi$ is
\be
{\cal L}_{Z'}= i{\bar\chi}\gamma^\mu D_\mu\chi- m_\chi {\bar\chi}\chi -\frac{1}{4}F'_{\mu\nu}F^{\prime \mu\nu}-\frac{1}{2}m^2_{Z'} Z'_{\mu}Z^{\prime \mu}+{\cal L}_{CS}
\ee
where the covariant derivative is $D_\mu \chi=(\partial_\mu-i g_{Z'} Z'_{\mu}/2)\chi$ and 
the effective $Z'$ interactions can come from the Chern-Simons(CS) terms as follows\footnote{The first two terms were considered in fermion dark matter models in Ref.~ \cite{zprimeCS}},
\be
{\cal L}_{CS}=a_1 \epsilon^{\mu\nu\rho\sigma} Z'_\mu Z_\nu F^Y_{\rho\sigma} +a_2 \epsilon^{\mu\nu\rho\sigma} Z'_\mu Z_\nu F'_{\rho\sigma}+b_1 \epsilon^{\mu\nu\rho\sigma} Z'_\mu (W^+_\nu \partial_\rho W^-_\sigma - \partial_\rho W^+_\nu W^-_\sigma).  \label{CSterm}
\ee
Here, $F'_{\rho\sigma}=\partial_\rho Z'_\sigma-\partial_\sigma Z'_\rho$ 
and $a_1, a_2, b_1$ are the dimensionless parameters generated by the triangle loops with extra heavy fermions.
We assume that there is no direct coupling between the $Z'$ gauge boson and the light quarks in the SM, apart from a small kinetic mixing with the $Z$-boson. Then, we can avoid the strong dijet bound and the electroweak precision constraint for the $Z'$ gauge boson. 

\begin{figure}[t]
\centering%
\includegraphics[width=7cm,natwidth=610,natheight=642]{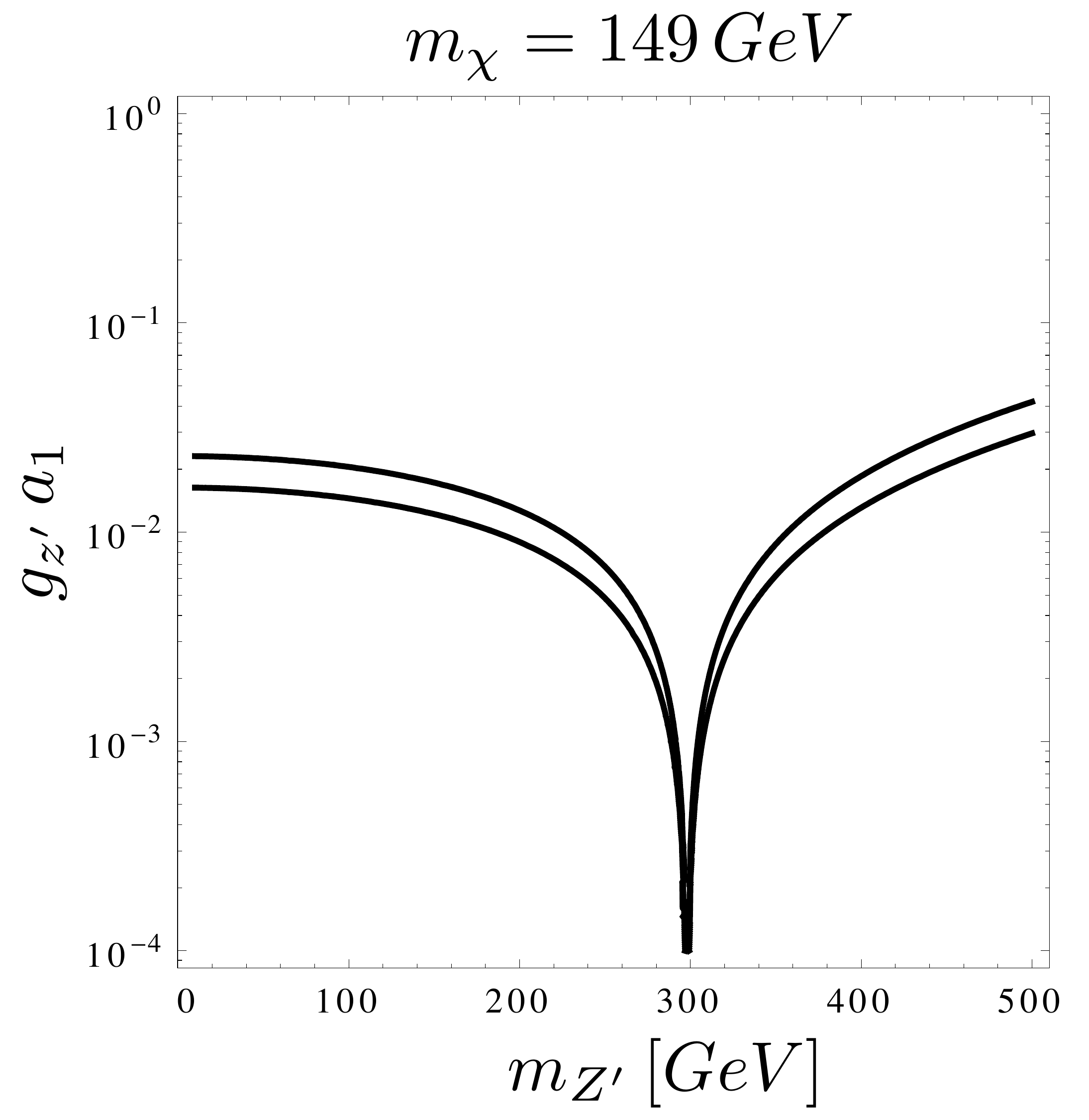}
\caption{ Region between solid lines is the parameter space of $m_{Z'}$ vs $g_{Z'}\,a_1$ satisfying $\langle\sigma v\rangle_{Z\gamma}/\langle \sigma v\rangle_0=0.08-0.16$ where $\langle\sigma v\rangle_0=3\times 10^{-26}\,{\rm cm}^3/{\rm s}$ is the thermal cross section. We have set dark matter mass to $m_\chi=149\,{\rm GeV}$ for Fermi gamma-ray line at $135\,{\rm GeV}$.
}\label{fig:Zprime}
\end{figure}

Furthermore, we can write the effective interaction between the $Z'$ gauge boson and the Higgs as
\be
\Delta{\cal L}_{CS}=\frac{a_3}{M} \, \epsilon^{\mu\nu\rho\sigma} Z'_\mu \partial_\nu  h\, F^Y_{\rho\sigma}
\label{CSHiggs}
\ee
where $a_3$ is a dimensionless parameter and $M$ is the cutoff scale.
As in the model in Ref.~\cite{zprimeTquark}, when top quark couples to the $Z'$ gauge boson, the Higgs coupling $\alpha_3$ can be generated by top quark running in the triangle loop.

The first coupling $a_1$ in eq.~(\ref{CSterm}) gives rise to the DM annihilation into $Z\gamma$, which can explain the Fermi gamma-ray line. Using the results in appendix A, we get the DM annihilation cross sections into $Z\gamma$ and $ZZ$, respectively, as
\bea
\langle\sigma v\rangle_{Z\gamma}&=&\frac{g^2_{Z'}a^2_1 \cos^2\theta_W}{32\pi} \frac{16 m^4_\chi}{(4m^2_\chi-m^2_{Z'})^2+\Gamma^2_{Z'}m^2_{Z'}}\,\left(1-\frac{m^2_Z}{4m^2_\chi}\right)^3 \Big(\frac{1}{m^2_Z}+\frac{1}{4m^2_\chi}\Big),  \label{annZg} \\
\langle\sigma v\rangle_{Z Z}&=&\frac{g^2_{Z'}a_1^2 \sin^2\theta_W}{32\pi m^2_Z} \frac{16 m^4_\chi}{(4m^2_\chi-m^2_{Z'})^2+\Gamma^2_{Z'}m^2_{Z'}}\,\left(1-\frac{m^2_Z}{m^2_\chi}\right)^{5/2}\label{annZZ} 
\eea
where the decay rate of the $Z'$ gauge boson is given by
\be
\Gamma_{Z'}=\Gamma_{Z'}(Z\gamma)+\Gamma_{Z'}(ZZ)
+\Gamma_{Z'}(WW)+\Gamma_{Z'}(\chi{\bar\chi})
\ee
with
\bea
\Gamma_{Z'}(Z\gamma)&=&\frac{m^3_{Z'}}{24\pi}\,a^2_1 \cos^2\theta_W\,\left(1-\frac{m^2_Z}{m^2_{Z'}}\right)^3 \Big(\frac{1}{m^2_Z}+\frac{1}{m^2_{Z'}}\Big),  \label{ZpZg} \\
\Gamma_{Z'}(ZZ)&=&\frac{m^3_{Z'}}{24\pi m^2_Z}\,a^2_1\sin^2\theta_W\, \Big(1-\frac{4m^2_Z}{m^2_{Z'}}\Big)^{5/2},  \label{ZpZZ}  \\
\Gamma_{Z'}(WW)&=& \frac{b^2_1 m^3_{Z'}}{48\pi m^2_W} \,\left(1-\frac{4m^2_W}{m^2_{Z'}}\right)^{5/2},    \label{ZpWW}   \\
\Gamma_{Z'}(\chi{\bar\chi})&=&\frac{g^2_{Z'}}{48\pi m_{Z'}}\,(m^2_{Z'}+2m^2_\chi) \left(1-\frac{4m^2_\chi}{m^2_{Z'}}\right)^{1/2}.  \label{Zpchichi}
\label{GammZp}
\eea
We note that when $m_{Z'}<m_\chi$, the $Z'$ gauge coupling to dark matter opens up the t-channel annihilation into $Z'Z'$, with the cross section,
\be
\langle \sigma v\rangle_{Z'Z'}=\frac{g^4_{Z'}}{64\pi}\frac{m^2_\chi}{(m^2_{Z'}-2m^2_\chi)^2}\Big(1-\frac{m^2_{Z'}}{m^2_\chi}\Big)^{3/2}.
\ee

Furthermore, there are extra DM annihilation channels coming from the other CS couplings in eq.~(\ref{CSterm}). First, for $m_{Z'}<2m_\chi-m_Z$, the  CS coupling $a_2$ contributes the s-channel annihilation into $Z' Z$ as 
\be 
\langle \sigma v\rangle_{Z'Z}=\frac{g^2_{Z'} a_2^2 \, m_\chi^2 \,  \lambda^{{\scriptscriptstyle1/2}}(4 ,R_{Z'}, R_Z)}{128\pi}
\left[\frac{\frac{\lambda(20,R_{Z'},R_Z)}{4}+ \frac{4 R_{Z'}+R_Z}{R_{Z'} R_Z}\,\lambda(4,R_{Z'}, R_Z)+4 (8 R_{Z'}-R_Z)}{(4m^2_\chi-m^2_{Z'})^2+\Gamma^2_{Z'}m^2_{Z'}} \right],\label{annZZp} 
\ee
where
\begin{eqnarray}
&&\lambda(x,y,z)= x^2+y^2+z^2-2 x y -2 y z -2 z x, \\
&&R_{Z'}= \left(\frac{m_{Z'}}{m_\chi}\right)^2, \quad R_Z= \left(\frac{m_{Z}}{m_\chi}\right)^2 \, .
\end{eqnarray}
Second, the CS coupling $b_1$ for the W-boson also leads to the s-channel annihilation into a $W^+W^-$ pair, with the cross section, 
\be
\langle\sigma v\rangle_{WW}=\frac{g^2_{Z'}b^2_1}{64\pi m^2_W} \frac{16 m^4_\chi}{(4m^2_\chi-m^2_{Z'})^2+\Gamma^2_{Z'}m^2_{Z'}}\,\left(1-\frac{m^2_W}{m^2_\chi}\right)^{5/2} \,.  \label{annWW}
\ee

\begin{figure}[t]
\centering%
\includegraphics[width=13cm,natwidth=610,natheight=642]{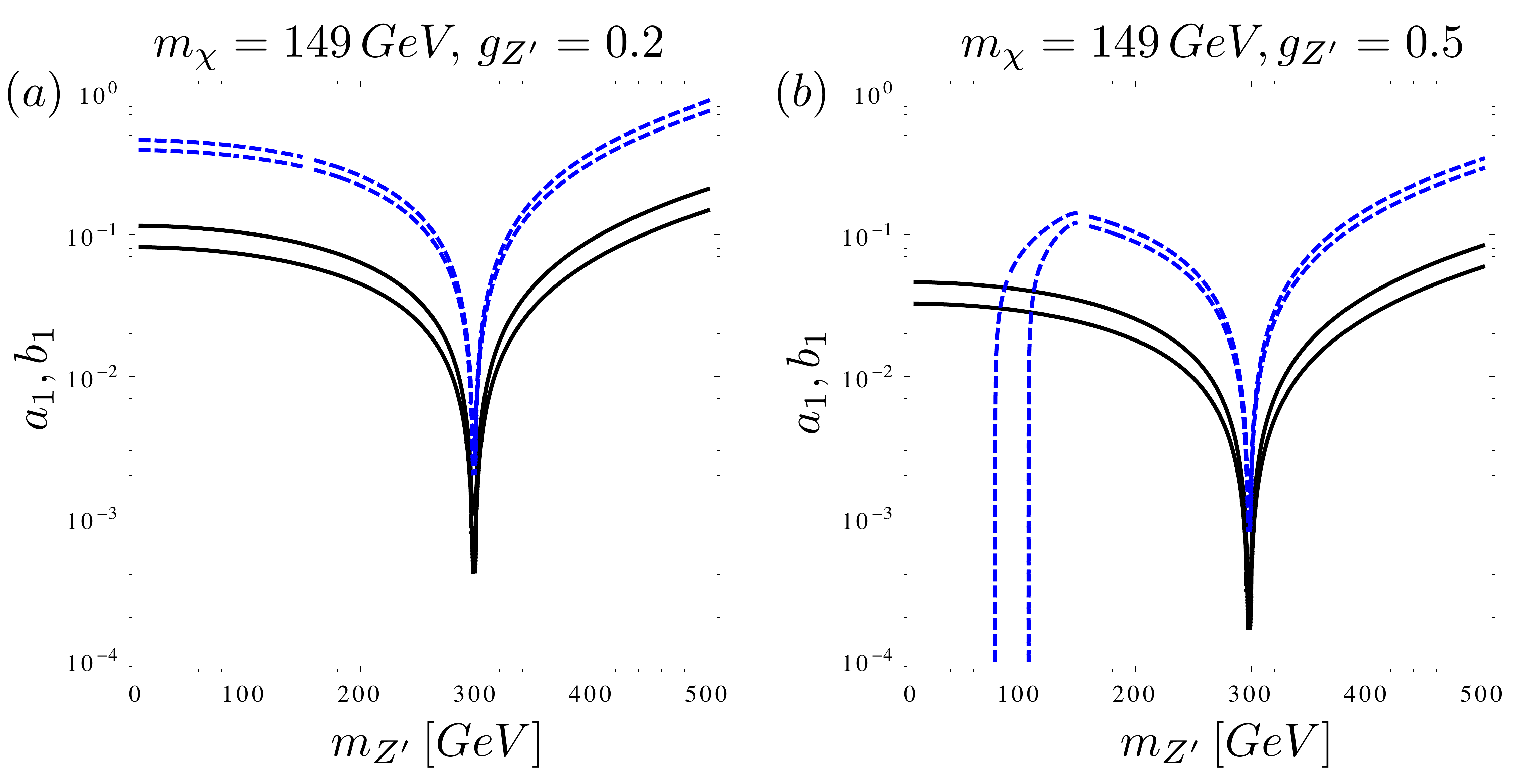}
\caption{ Region between black solid lines: parameter space of $m_{Z'}$ vs $a_1$ satisfying $\langle\sigma v\rangle_{Z\gamma}/\langle \sigma v\rangle_0=0.08-0.16$ where $\langle\sigma v\rangle_0=3\times 10^{-26}\,{\rm cm}^3/{\rm s}$ is the thermal cross section. Region between blue dashed lines: parameter space of $m_{Z'}$ vs $b_1$ producing the relic density within WMAP $5\sigma$. We have chosen $g_{Z'}=0.2$ (Left), $0.5$ (Right)
and set DM mass to $m_\chi=149\,{\rm GeV}$ for Fermi gamma-ray line at $135\,{\rm GeV}$.
}\label{fig:Zprime2}
\end{figure}

In $Z'$-mediated dark matter models, there is no DM annihilation into a photon pair due to Landau-Yang theorem. Instead, the DM annihilation into $Z\gamma$ or $h\gamma$ leads to a gamma-ray line at $E_\gamma=m_\chi(1-m^2_{Z,h}/(4m^2_\chi))$. Thus, in order to obtain the Fermi gamma-ray line at $130(135)\,{\rm GeV}$, we need to take $m_\chi=144(149)\,{\rm GeV}$  for $Z\gamma$ or $m_\chi=155(160)\,{\rm GeV}$ for $h\gamma$ where we chose the Higgs mass, $m_h=126\,{\rm GeV}$. In Fig.~\ref{fig:Zprime}, we depict the parameter space compatible with the DM annihilation cross section for $\chi{\bar\chi}\rightarrow Z\gamma$, on the $m_{Z'}$ vs $g_{Z'}a_1$ plane for $m_{\chi}=149\,{\rm GeV}$. In order to get the correct photon flux, we need the DM annihilation cross section into a single photon to be twice the fitted value of the DM annihilation cross section into a photon pair as discussed in the previous section.

Now we discuss the constraint of the relic density on the model.
First,  if $m_\chi< m_{Z'}<2m_\chi-m_Z$, there is an extra channel with $Z' Z$ coming from the $\alpha_2$ coupling, which allows the possibility to obtain a thermal cross section. In this case, the second coupling $a_2$ in eq.~(\ref{CSterm}) is constrained to satisfy the WMAP constraint.
Second, if $m_{Z'}< m_\chi$,  the t-channel DM exchange leads to the main contribution to dark matter annihilation so the relic density condition is independent of $a_2$. 
As a result, the relic density  condition can be satisfied by the $Z'$ gauge coupling or the $a_2$ coupling, in the low $Z'$ mass region below $2m_\chi- m_Z$.
Lastly, for $m_{Z'}>2 m_\chi -m_Z$, there are two final states with $ZZ$ and $Z\gamma$ coming from the $a_1$ coupling. But, since $\langle\sigma v\rangle_{ZZ}/\langle\sigma v\rangle_{Z\gamma}\sim 0.3$, the $Z\gamma$ final state is always dominant. In this case, we could not get the correct relic density without extra channels.
We note that if kinematically allowed, the CS coupling of the W-boson in eq.~(\ref{CSterm}) can open an extra annihilation channel with a $W^+W^-$ final state. Then, it is possible to explain the relic density with the $b_1$ coupling only, even for $m_{Z'}> 2m_\chi-m_Z$. In the following discussion, we focus on the CS couplings $a_1$ and $b_1$, in particular, in the high $Z'$ mass region.

We overlay in Fig.~\ref{fig:Zprime2} the parameter for $m_{Z'}$ vs $b_1$ satisfying the WMAP bound on the relic density within $5\sigma$, in comparison to the parameter space of the $a_1$ coupling shown in Fig.~\ref{fig:Zprime}. In the low $Z'$ mass region below $m_\chi$, the contribution of the $WW$ channel to thermal cross section can be bounded due to the $Z'Z'$ channel, depending on the value of the $Z'$ gauge coupling. On the other hand, for $m_{Z'}> m_\chi$, the thermal cross section is determined dominantly by the $WW$ channel, so it only depends on the product of the couplings, $g_{Z'}b_1$. We find that when the $Z'Z'$ channel is subdominant, the CS coupling for the $W$-boson should be larger than the one for the $Z$-boson by order of magnitude. 
We note that the continuum photons coming from the WW channel and other channels without monophotons are consistent with the gamma-ray \cite{dwarfgalaxy} and anti-proton \cite{antiproton} constraints  and can be also compatible with the shape of the line spectrum \cite{continuum}.

Now we comment on the case that the top quark couples to the $Z'$ gauge boson with the same $Z'$ charge as dark matter. In this case, the same $Z'$ gauge coupling is shown to generate the CS coupling to the Higgs boson  \cite{zprimeTquark}, as given in eq.~(\ref{CSHiggs}).
From the $a_3$ coupling in eq.~(\ref{CSHiggs}), we obtain the annihilation cross section for $\chi{\bar\chi}\rightarrow h\gamma$ as follows,
\be
\langle\sigma v\rangle_{h\gamma}=\frac{g^2_{Z'}a^2_3\cos^2\theta_W}{32\pi M^2}\, \frac{16m^4_\chi}{(4m^2_\chi-m^2_Z)^2+\Gamma^2_{Z'}m^2_{Z'}}\, \bigg(1-\frac{m^2_h}{4m^2_\chi}\bigg)^3. \label{annhg}
\ee
So, the DM annihilation into $h\gamma$ can give rise to an extra line at the energy smaller than the photon energy at the Fermi line. From eq.~(\ref{annZg}), the annihilation cross section for $\chi{\bar\chi}\rightarrow h\gamma$ is comparable to the one when $\chi{\bar\chi}\rightarrow Z\gamma$ for $M/a_3\simeq m_Z/a_1$. In this case, we can get two gamma-ray lines at $E_\gamma=116(122)$ and $130(135)\,{\rm GeV}$ for $m_\chi=144(149)\,{\rm GeV}$ from $h\gamma$ and $Z\gamma$, respectively.
Furthermore, if kinematically allowed, there are extra decay modes of the $Z'$-boson as follows,
\bea
\Gamma_{Z'}(h\gamma)&=& \frac{m^3_{Z'}}{24\pi M^2}\,a^2_3\cos^2\theta_W \Big(1-\frac{m^2_h}{m^2_{Z'}}\Big)^3, \\
\Gamma_{Z'}(hZ)&=& \frac{m^3_{Z'}}{24\pi M^2}\,a^2_3\sin^2\theta_W \Big(1-\frac{(m_h+m_Z)^2}{m^2_{Z'}}\Big)^{3/2} \Big(1-\frac{(m_h-m_Z)^2}{m^2_{Z'}}\Big)^{3/2}, \\
\Gamma_{Z'}( t{\bar t})&=& \frac{N_c g^2_{Z'}}{48\pi m_{Z'}} (m^2_{Z'}+2 m^2_t) \Big(1-\frac{4m^2_t}{m^2_{Z'}}\Big)^{1/2}.
\eea
On the other hand, the top quark coupling to $Z'$ can open a new DM annihilation channel into a $t{\bar t}$ final state and allow for $Z'$-boson to be heavier than $2m_\chi-m_Z$, in which case the $ZZ$ and $Z\gamma$ annihilation channels only cannot explain the relic density.
Suppose that the $W$-boson CS coupling is small. Then, for $m_{Z'}\gtrsim 200\,{\rm GeV}$ and $m_\chi \lesssim m_t$, the relic density can be obtained by the top quark coupling to $Z'$, when $m_\chi$ is slightly above or below $m_{Z'}/2$, i.e. close to the $Z'$ resonance, or $Z'$ couplings are large, for which $m_\chi$ slightly smaller than $m_t$ is preferred.  For $m_\chi\gtrsim m_t$, the DM annihilation proceeds dominantly into a $t{\bar t}$ final state, so, for moderate couplings, $g_{Z'}\sim 1$, there is a robust prediction for the dark matter mass, whatever the value $m_{Z'}\gtrsim 350\,{\rm GeV}$: $m_\chi\sim 150\,{\rm GeV}$, which is far away from the $Z'$ resonance \cite{zprimeTquark}. In this case, we note that the $Z'$-boson can also decay into a $t{\bar t}$ pair with the branching fraction larger than the one of the $\chi{\bar\chi}$ decay mode.

\subsection{Summary}

In axion mediation, Fermi gamma-ray line at $130(135)\,{\rm GeV}$ is explained mostly by the DM annihilation into a photon pair and it requires $m_\chi=130(135)\,{\rm GeV}$ and favors $c_2/c_1 >1$ for the correct branching fraction of the corresponding annihilation cross section. Since we rely on the axion resonance for the enhancement of the annihilation cross section, the axion mass range is typically constrained to be close to $m_a=260(270)\,{\rm GeV}$ for $c_1, c_2$ of order one.  Varying $v_s$ (on the left of Fig.~\ref{fig:axion}) and/or $c_2/c_1$ values (on the right of  Fig.~\ref{fig:axion}), we may relax the axion mass up to as large as $500\,{\rm GeV}$ for $v_s\sim 55\,{\rm GeV}$ and $(c_1,c_2)\sim (1,10)$.

In $Z'$ mediation, the DM annihilation into $Z\gamma$ or $h\gamma$ can explain Fermi gamma-ray line at $130(135)\,{\rm GeV}$ for $m_\chi=144(149)\,{\rm GeV}$ or $155(160)\,{\rm GeV}$. The correct value of the corresponding annihilation cross section can be obtained for $g_{Z'} a_1\sim 0.01$ or $g_{Z'} a_3\sim 0.01 M /m_Z$, far away from the $Z'$ resonance. 
If there is a $W$-boson CS coupling, there is no limit on the $Z'$-boson mass coming from the relic density, because we can vary the $W$-boson CS coupling to satisfy the relic density condition. In this case, the relic density is obtained by the $WW$ annihilation channel with $g_{Z'} b_1\sim 0.1$, far away from the $Z'$ resonance.
If there is no $W$-boson coupling, the relic density condition requires $m_{Z'}< 2m_\chi-m_Z$ in the absence of the top coupling to $Z'$ and $m_{Z'}\gtrsim 200\,{\rm GeV}$ in the presence of the top coupling to $Z'$.
The top quark coupling to $Z'$ would open up an extra $Z'$ decay mode into $t{\bar t}$ when $m_{Z'}>2 m_\chi$.
In order to maximize the monophoton events with large missing energy in the next section, henceforth, we assume the case with a sizable $W$-boson CS coupling and no or small top gauge coupling to $Z'$.

%%%%%%%%%%%%%%%%%%%%%%%%%%%%
\section{Mediator production at colliders}
%%%%%%%%%%%%%%%%%%%%%%%%%%%

\begin{figure}[t]
\centering%
\includegraphics[width=15.5cm,natwidth=610,natheight=642]{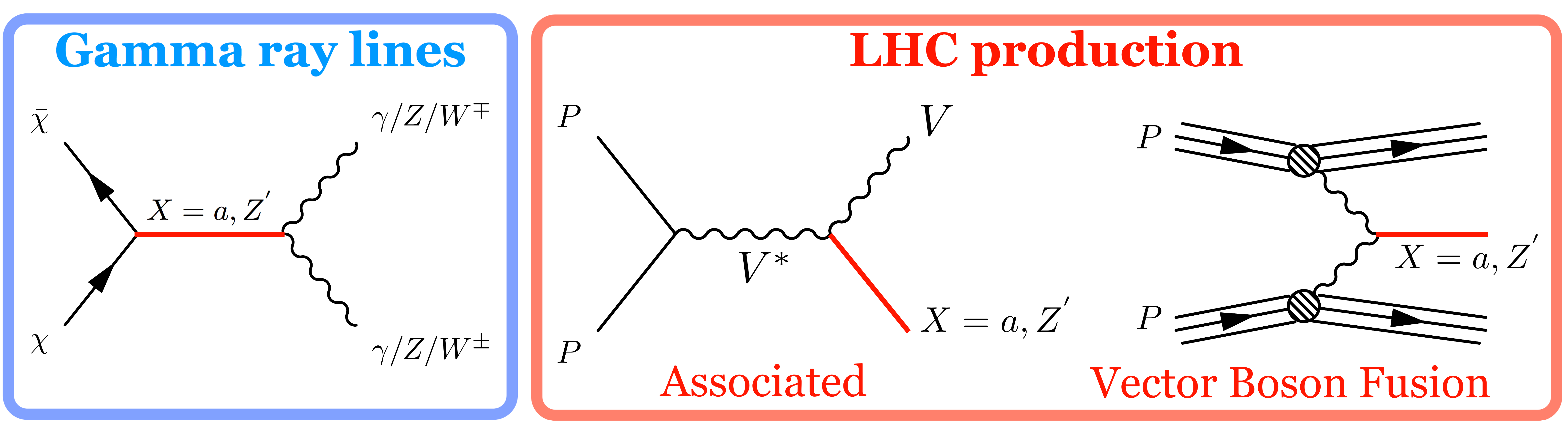}
\caption{The relation between lines in the gamma spectrum and possible signatures at the LHC.}
\label{diagrams}
\end{figure}

In this section we consider the production of the axion and $Z'$ mediators at the LHC. 
The same diagram which mediates dark matter annihilation into two vector bosons, can be rotated to lead to LHC production of the mediator, see Fig.~\ref{diagrams}.

Unlike the approaches with the effective interactions between dark matter and the SM quarks  \footnote{The collider searches of a monophoton coming from the initial state radiation have been considered in Ref.~\cite{monogsearches}.}, the dominant production is in association with a vector boson
\bea
p \, p \to V^* \to V + X
\eea
where $X=a,Z'$ and $V=\gamma$, $Z$ or $W^{\pm}$. One could also produce the mediator in vector boson fusion (VBF) processes 
\bea
p \, p \to X+ j \, j
\eea
where the jets $j$ are rather forward~\cite{VBF}. We will discuss this mode in Sec.~\ref{14tev}.

\subsection{Axion production}

In axion-mediated dark matter models, the same interactions producing the DM annihilation into a photon pair can lead to monophoton events in association with a dark matter pair, $pp\rightarrow \gamma\, a$, at the LHC.

When $m_a>2m_{\chi}$ in the half region of the axion resonance, the axion can decay into a dark matter pair dominantly for the dark matter coupling $\lambda_\chi$ of order one, see eqs.~\ref{GammAx}. If $m_a<2m_\chi$ in the other half region of the axion resonance, the dark matter production is kinematically suppressed so the production of a pair of electroweak gauge bosons is dominant.  In the case where the axion decays into $ZZ$ or $Z\gamma$, one would obtain missing energy from the decay of a Z-boson into a neutrino pair. 
Here, we assume that the extra vector-like lepton running in the loops is heavy enough as $m_a<2m_f$ such that the  extra vector-like charged lepton pair is not produced from the axion decay.

For the monophoton production at the LHC, the only relevant parameters are the axion decay constant $v_s$, the anomaly coefficients, $c_{1,2}$, the axion mass $m_a$ and DM mass $m_\chi$.
The parton-level production cross section for $q{\bar q}\rightarrow a\,\gamma$ is \footnote{See Appendix~B for details.}
\bea
\sigma(q{\bar q}\rightarrow a\,\gamma)&=& \frac{2\alpha}{3N_c}\bigg[ 4 Q^2_q c^2_{\gamma\gamma}+
\frac{(v^2_q+a^2_q) c^2_{Z\gamma}}{\sin^2(2\theta_W)}\, \frac{s^2}{(s-m^2_Z)^2+\Gamma^2_Z m^2_Z} \nonumber \\
&&\quad+ \frac{4 Q_q v_q c_{\gamma\gamma} c_{Z\gamma}}{\sin(2\theta_W)}\,\frac{s(s-m^2_Z)}{(s-m^2_Z)^2+\Gamma^2_Z m^2_Z} \bigg] \Big(1-\frac{m^2_a}{s}\Big)^3
\eea
where $v_q=T^q_3(1-4|Q_q|\sin^2\theta_W)$ and $a_q=T^q_3$, that is, 
\be
v_u=\frac{1}{2}\Big(1-\frac{8}{3}\sin^2\theta_W\Big), \,\, v_d=\frac{1}{2}\Big(-1+\frac{4}{3}\sin^2\theta_W\Big), \,\, a_u=-a_d=\frac{1}{2}.
\ee
Similarly, the production cross section for $q{\bar q}\rightarrow a\, Z$ is
\bea
\sigma(q{\bar q}\rightarrow a\,Z)&=& \frac{2\alpha}{3N_c}\bigg[ Q^2_q c^2_{Z\gamma}+
\frac{4 (v^2_q+a^2_q) c^2_{ZZ}}{\sin^2(2\theta_W)}\, \frac{s^2}{(s-m^2_Z)^2+\Gamma^2_Z m^2_Z} \nonumber \\
&&\quad+ \frac{4 Q_q v_q c_{Z\gamma} c_{ZZ}}{\sin(2\theta_W)}\,\frac{s(s-m^2_Z)}{(s-m^2_Z)^2+\Gamma^2_Z m^2_Z} \bigg] 
\lambda^{\frac{3}{2}}{ \left(1, \frac{m_a^2}{s},\frac{m_Z^2}{s}\right)}\, .
\eea

%*******************************
\subsection{$Z'$ production}
%*******************************

In $Z'$-mediated dark matter models, due to the same CS coupling leading to the DM annihilation into $Z\gamma$,  there is a similar process, $q{\bar q}
\rightarrow \gamma Z'$, where off-shell $Z$-boson produces a $Z'$ gauge boson in association with a monophoton.

For  $m_{Z'}>2m_\chi$,  $Z'$-boson can decay into a DM pair, producing a large missing energy.
In order for the branching fraction into $\chi{\bar\chi}$ to be dominant, we need $g_{Z'}\gtrsim a_1, b_1$, from eqs.~(\ref{ZpZg})-(\ref{Zpchichi}). Thus, from the conditions for Fermi  gamma-ray line and relic density obtained in the previous section, $g_{Z'} a_1\sim 0.01$ and $g_{Z'} b_1\sim 0.1$,  we get the bounds on the CS couplings, $a_1\lesssim 0.1$ and $b_1\lesssim 0.3$, for the dominance of the $\chi{\bar\chi}$ decay mode.
Below the mass threshold of a DM pair, $Z'$-boson decays into $Z\gamma$ and/or $ZZ, WW$. In the case where the $Z'$-boson decays into $Z\gamma$, we have two photons and  missing energy coming from the decay of a Z-boson into a neutrino pair. When the $Z'$-boson decays into $ZZ$, we have monophoton events with missing energy coming from four neutrinos, but the $ZZ$ decay mode is suppressed by $\sin^2\theta_W$, see eqs.~(\ref{GammZp}).

For the monophoton production at the LHC, the relevant parameters are the $Z'$ CS coupling $a_1$ and  the $Z'$ mass $m_{Z'}$.
The parton-level production cross section for $q{\bar q}\rightarrow Z'\,\gamma $ is
\bea
\sigma(q{\bar q}\rightarrow Z'\,\gamma)&=&\frac{1}{96\pi N_c}\frac{g^2 a_1^2  (v^2_q+a^2_q)}{(s-m^2_Z)^2+m^2_Z\Gamma^2_Z}\, s \bigg( 1+\frac{s}{m^2_{Z'}}\bigg) \bigg(1-\frac{m^2_{Z'}}{s}  \bigg)^3. 
\eea
See Appendix~B for details.

%********************************
\section{Setting Limits}
%******************************

We extract the trivial dependence on couplings by defining a re-scaled cross section
\bea
\tilde{\sigma} = \frac{\sigma}{c^2}
\eea
where 
\bea
c = \frac{c_1+c_2}{v_s/v} \textrm{ for the axion, and } a_1 \textrm{ for the } Z' \ ,
\eea
with $v=246$ GeV.

\begin{figure}[t]
\centering%
\includegraphics[width=9cm,natwidth=610,natheight=642]{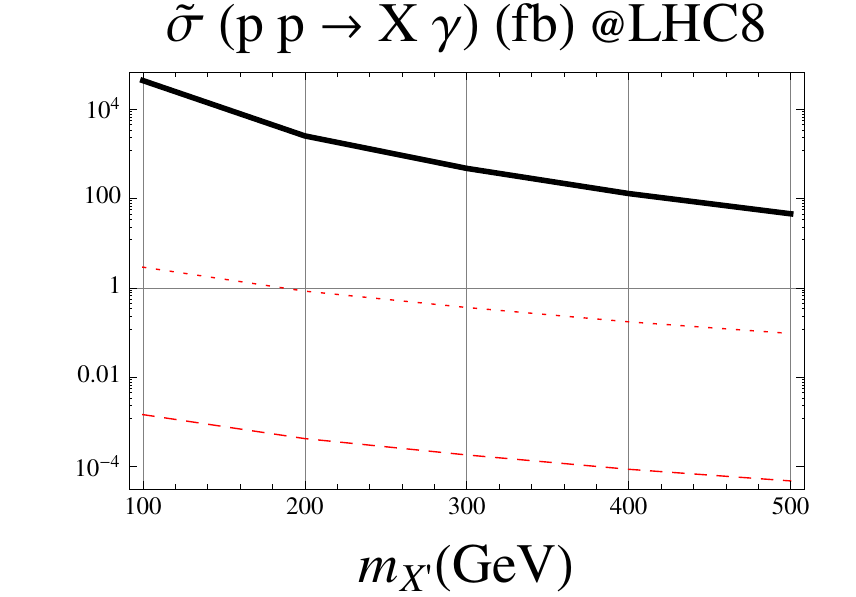}
\caption{Parton-level production cross section for $q{\bar q}\rightarrow (a, Z')+ \gamma$ at LHC $8\,{\rm TeV}$. Red-dashed line corresponds to the axion case, black-solid is the $Z'$ case. For illustration, we show the axion red-dotted line with $c_{1,2}=10$ and $v_s=55$ GeV.}
\label{assoprod}
\end{figure}

For illustration, we show in Fig.~\ref{assoprod} the scaled cross section $\tilde\sigma$ in association with a photon. The axion cross section is much smaller 
because in the coefficients $c_i$ one has already pulled out the $\alpha_i/8 \pi$ prefactor, plus the axion coupling have one more derivative. This extra-derivative makes the off-shell vector boson to have a larger invariant mass~\cite{MVX}, resulting in a PDF suppression. Nevertheless, if $v_s$ is small and $c_{1,2}$ relatively large, the model can be accessible, see the red-dotted line in Fig.~\ref{assoprod}.

In setting limits, we focus on the signature of a photon recoiling against missing energy. Both CMS~\cite{CMS-monoa} and ATLAS~\cite{ATLAS-monoa} carry this search, where the selection criteria is as follows
\bea
& & \textrm{ 1. Isolated photon with } |\eta_{\gamma}|< 1.44 (2.37) \textrm{ and } p_T^{\gamma}> 145 (150) \textrm{ GeV} \\
& & \textrm{ 2. Missing Energy } \slashed{E}_T> 130 (150) \textrm{ GeV}
\eea
where the numbers in parenthesis correspond to ATLAS.
There are other isolation criteria, such as the requirement that the missing energy and hadronic activity, or the photon, should not be too close. Veto on a number of high-$p_T$ jets or leptons is also applied, but there is no veto on extra photons.  For details, look at the experimental notes.

Although the public notes correspond to the 7 TeV run in 2011, one can estimate the reach of the 2012 run at 8 TeV as shown in Ref.~\cite{monoa-vero}. Assuming we combine CMS and ATLAS, we expect a limit on the cross section
\bea
(\sigma \times BR \times Acc) \lesssim 1 \textrm{fb} \ 
\eea
where $BR$ is the branching fraction of a final state and $Acc$ is the acceptance.

In the following we are going to first discuss the axion model for the LEP bound and the 8 TeV run  (Sec.~\ref{8tevaxion}) and next the $Z'$ model for the 8 TeV run (Sec.~\ref{8tev}). We also discuss the expected reach to search for axion and $Z'$ in the 14 TeV upgrade (Sec.~\ref{14tev}).

%*************************
\subsection{Bounds on the axion from the 8 TeV run} \label{8tevaxion}

%*************************

There are three ways in which the axion is sensitive to the monophoton search. Above the DM threshold, we will consider the process
\bea
& & p \, p \to \gamma^*/Z^* \to \gamma \,a \  \nonumber \\
& & \textrm{ with } a \to \chi \tilde{\chi}\,.   \nonumber 
\eea
Below the threshold, we will rely on the processes
\bea
& & p \, p \to \gamma^*/Z^* \to Z Z \gamma \textrm{ or } Z \gamma \gamma \nonumber \\
 & & \textrm{ with at least one } Z\to \nu \bar{\nu}\,,  \nonumber 
\eea
\bea
& & p \, p \to W^* \to  Z \gamma \,W    \nonumber \\
 & & \textrm{ with } Z\to \nu \bar{\nu} \textrm{ and/or } W\to {\bar \nu} l\,. \nonumber 
\eea

\begin{figure}[t]
\centering%
\includegraphics[width=7.7cm,natwidth=610,natheight=642]{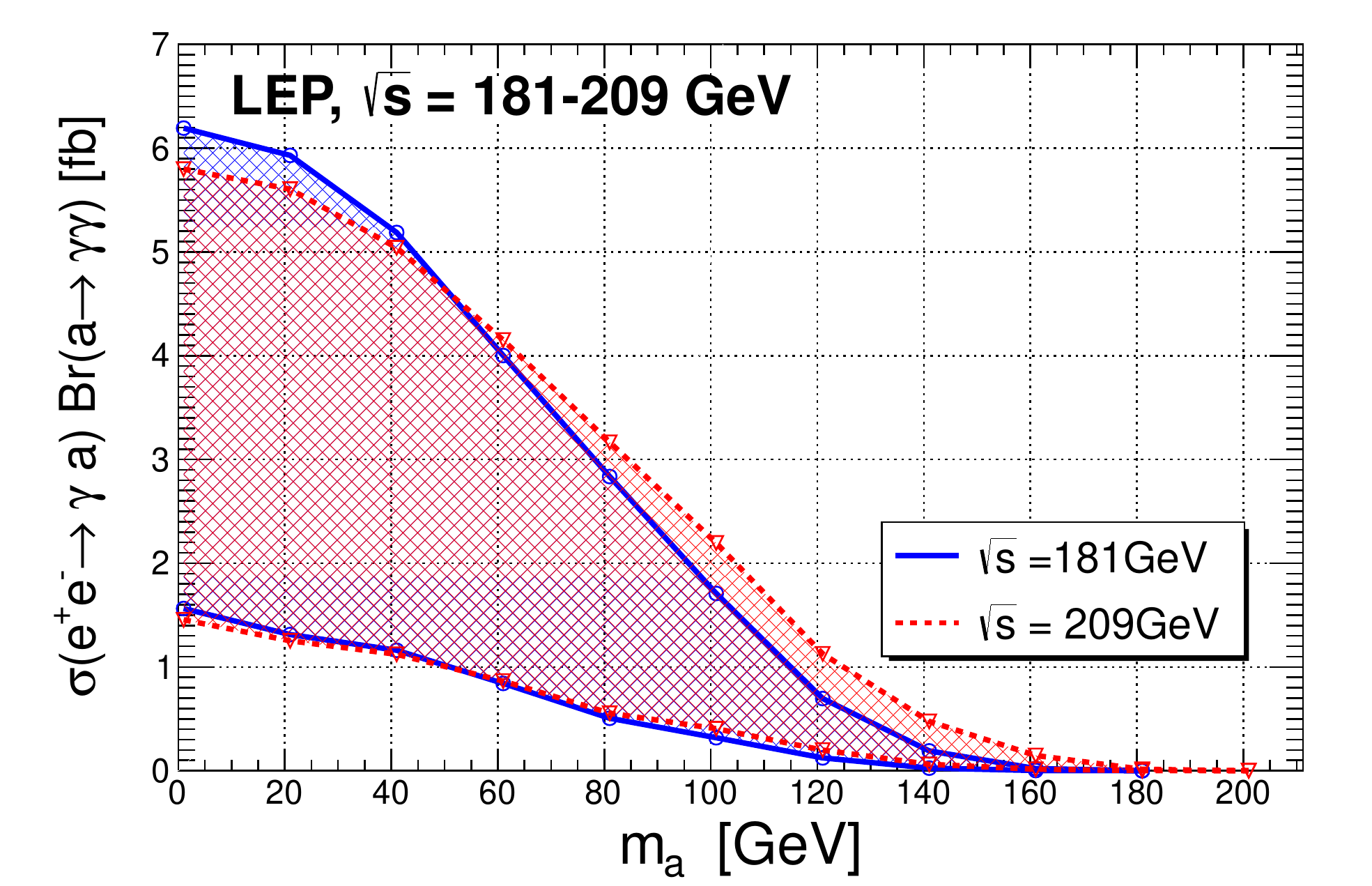}
\includegraphics[width=7.7cm,natwidth=610,natheight=642]{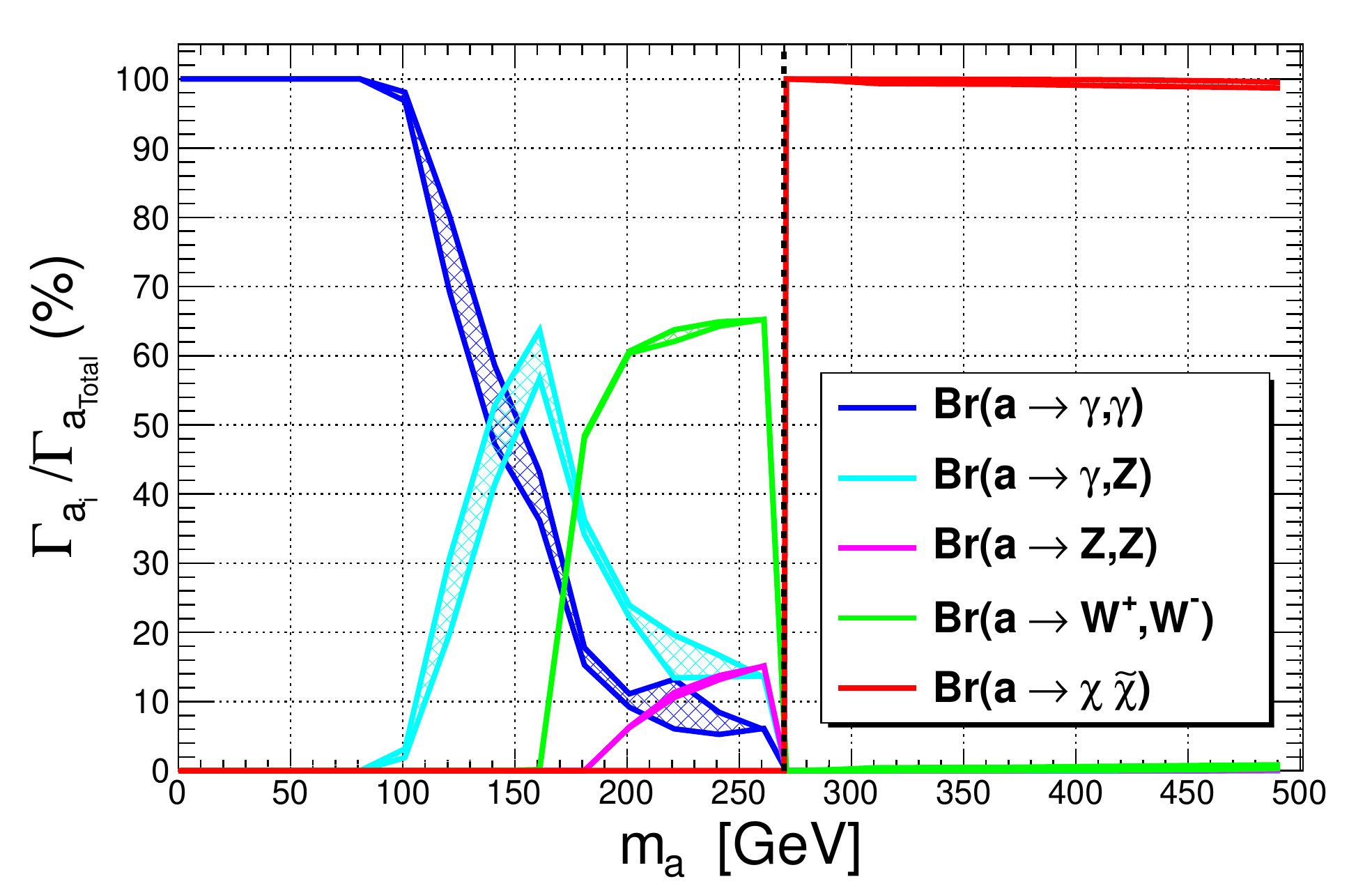}
\includegraphics[width=7.5cm,natwidth=610,natheight=642]{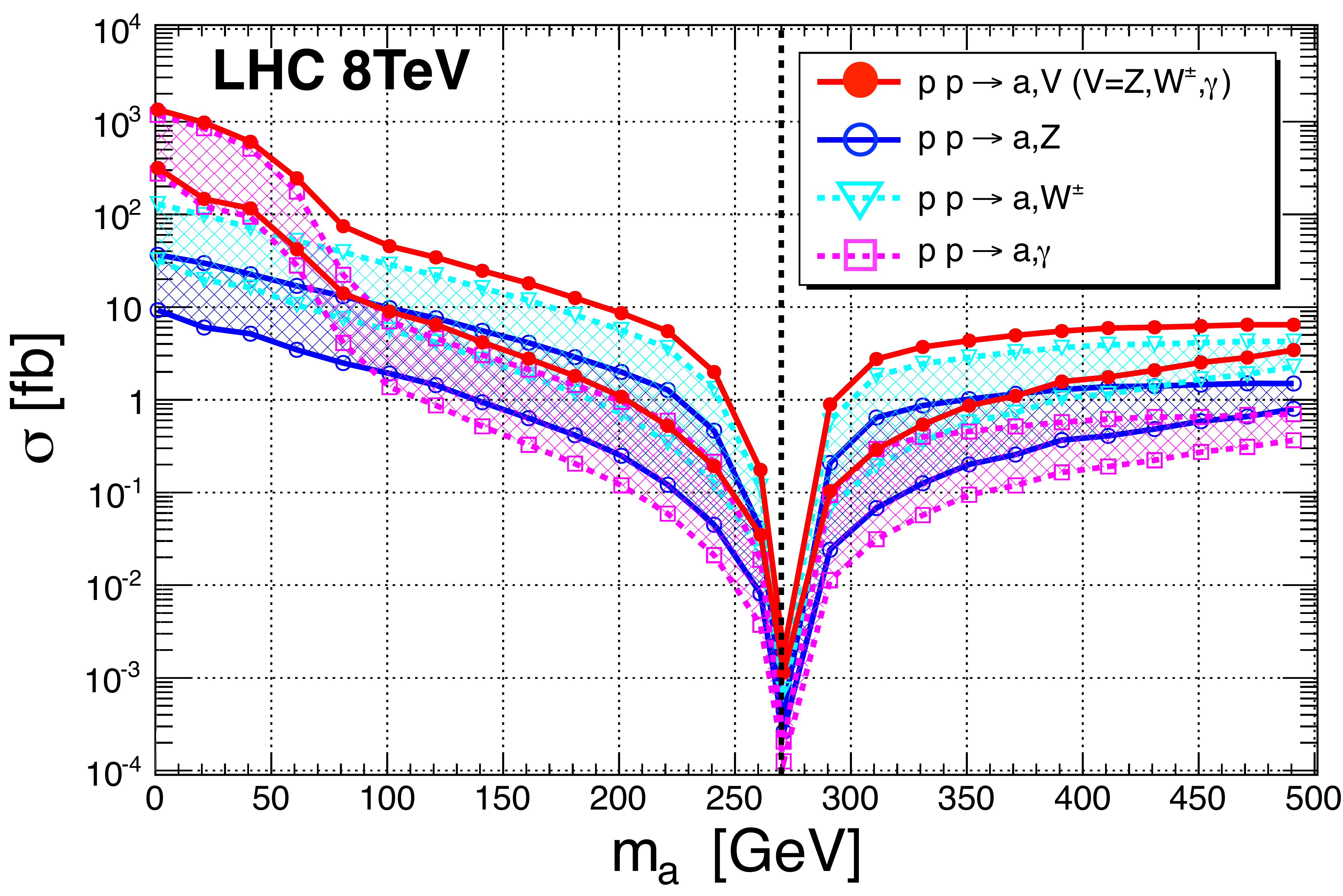}
\includegraphics[width=7.9cm,natwidth=610,natheight=642]{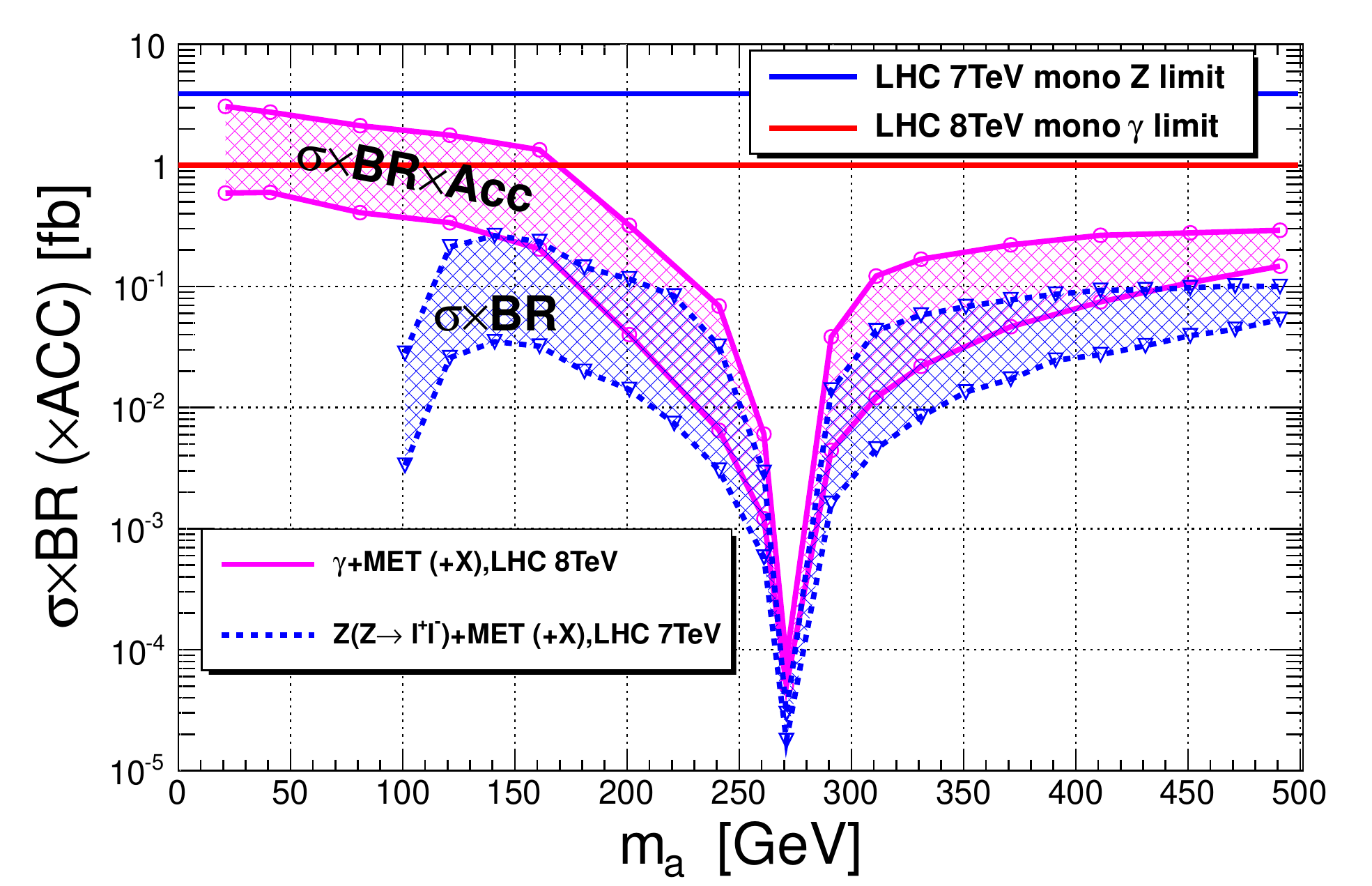}
\caption{Upper panel: (Left) Production cross section $e^+ e^-\rightarrow \gamma \,a\rightarrow \gamma\gamma\gamma$ at LEP, $\sqrt{s}=181-209\,{\rm GeV}$.  
(Right) Branching fraction of the axion decays below $200\,{\rm GeV}$.
Lower panel: (Left) Cross sections of associated production $p,p\to a,\,V$ at the LHC $8$ TeV in axion mediation. (Right) Production cross sections multiplied by branching fraction for $\gamma+{\rm MET}+X$ (magenta) at the LHC $8$ TeV and $Z(\rightarrow l^+ l^-)+{\rm MET}+X$ (blue) at the LHC 7 TeV and the limits for the corresponding processes (red solid and dashed lines, respectively).
DM mass is set to $m_\chi=135GeV$ and parameters $(v_s, c_1, c_2)$ are chosen to satisfy the Fermi gamma ray ratio and WMAP within $5\sigma$ from Fig~\ref{fig:axion}. Input parameters are chosen to give the maximum and minimum cross section for each axion mass, denoted as the boundary lines of the production cross sections.
}\label{fig:production}
\end{figure}

As shown in Fig.~\ref{assoprod}, the axion case would be difficult to access with the 8 TeV run,  unless $v_s \ll v$ and $c_{1,2}$ are order one. From the point of view of UV models, $v_s$ is given by the scale of the PQ symmetry breaking with a complex scalar $S$ containing the weak-scale axion \cite{axion2}, and it is related to the mass of a heavy fermion coupled to $S$ and the mass of the real partner of $S$ as
\bea
v_s =\sqrt{2}\, \frac{m_f}{\lambda_f}=\frac{m_s}{\sqrt{2\lambda_s}} 
\eea
with $\lambda_f$ being the Yukawa coupling of the heavy fermion to $S$ and $\lambda_s$ being the singlet quartic coupling. In order for the effective theory for dark matter and axion to make sense, we need to impose $m_f, m_s\gtrsim m_a\sim 270\,{\rm GeV}$, so we get $v_s\gtrsim 54\,{\rm GeV}$ for $\lambda_s=4\pi$.
One can use this bound, and other constraints to obtain a bound on $m_a$ as follows:

\bea
& & v_s \gtrsim 54 \textrm{ GeV, } c_{1,2} \lesssim 10 \textrm{ , } m_{\chi} \simeq 135   \textrm{ GeV,  and } \langle \sigma_{\gamma \gamma} v\rangle \simeq 10^{-27} {\rm cm}^3 {\rm s}^{-1}  \nonumber \\
& &  \Rightarrow m_{a} \lesssim 500  \textrm{ GeV}.  \nonumber 
\eea

In Fig.~\ref{fig:production}, we show in the upper panel  the multiphoton production cross section at LEP with $\sqrt{s}=181-209\,{\rm GeV}$ and the branching fraction of the axion decays  in the low axion mass region and also present in the lower panel the reaches from monophoton limit at the LHC 8 TeV and from mono-Z limit at the LHC 7 TeV \cite{LHCmonoZ,mono-Z}. The cross sections for all the associated axion productions are computed for parameters $(v_s, c_1, c_2)$ in the ranges of $55\,{\rm GeV}\leq v_s\leq 200\,{\rm GeV}$ and $0<c_{1,2}\leq10$, satisfying the Fermi gamma ray ratio and WMAP within $5\sigma$. As seen in the left of the upper panel, the axion production cross section with the axion decaying into a photon pair is below $6.2\,{\rm fb}$ at LEP with $\sqrt{s}=181-209\,{\rm GeV}$, which is within the experimental uncertainties of the measured cross section around $7-9.5\,{\rm pb}$ at LEP\cite{LEP3gamma}.
On the other hand, from the right plot in the lower panel of Fig.~\ref{fig:production}, we find that the region of the parameter space below $m_a=170\,{\rm GeV}$ can be constrained by current monophoton searches  but the axion production cross sections are below the current sensitivities for mono-Z limits.

Finally, we note that astrophysics also constrains the axion-photon coupling and the small axion masses.  If axions are produced inside a star and escape, they provide an additional cooling channel. Then, the non-standard energy loss prolongs the red giant  phase and shortens the horizon branch phase.  In our case, the bound of star cooling on the axion-photon coupling can be satisfied even for the weak-scale $v_s$, if the axion mass is $m_a\gtrsim 300\,{\rm keV}$ \cite{HB}.

%*************************
\subsection{Bounds on $Z'$ from the 8 TeV run} \label{8tev}

%*************************

In this section we focus on the $Z'$ models to discuss the reach of the 8 TeV run. 
There are two ways in which the $Z'$ is sensitive to the monophoton search. Above the DM threshold, we will consider the process
\bea
& & p \, p \to Z^* \to \gamma Z'   \nonumber \\
& & \textrm{ with } Z' \to \chi \tilde{\chi}\,.   \nonumber 
\eea
Below the threshold, we will rely on the processes
\bea
& & p \, p \to Z^* \to Z Z \gamma \textrm{ or } Z \gamma \gamma  \nonumber  \\  
 & & \textrm{ with at least one } Z\to \nu \bar{\nu}\,.  \nonumber
\eea
We note that in the presence of the $W$-boson CS term,
there is an extra process for the $Z'$ production in association with the $W$-boson and the $Z'$-boson can decay into a $WW$ pair. In this case, the produced $W$-boson(s) can also provide missing energy through the leptonic decay.

\begin{figure}[t]
\centering%
\includegraphics[width=7.5cm,natwidth=610,natheight=642]{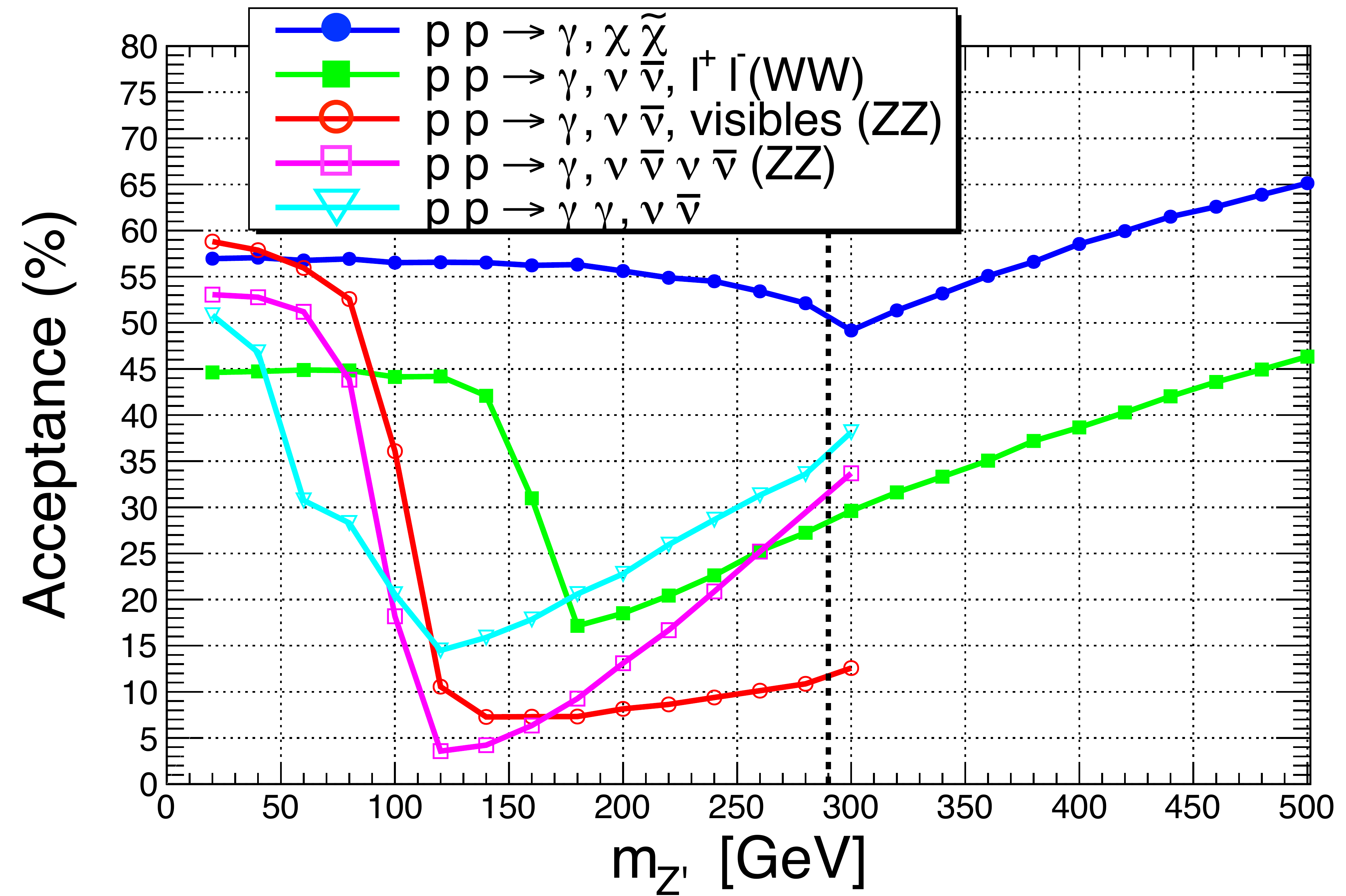}
\includegraphics[width=7.5cm,natwidth=610,natheight=642]{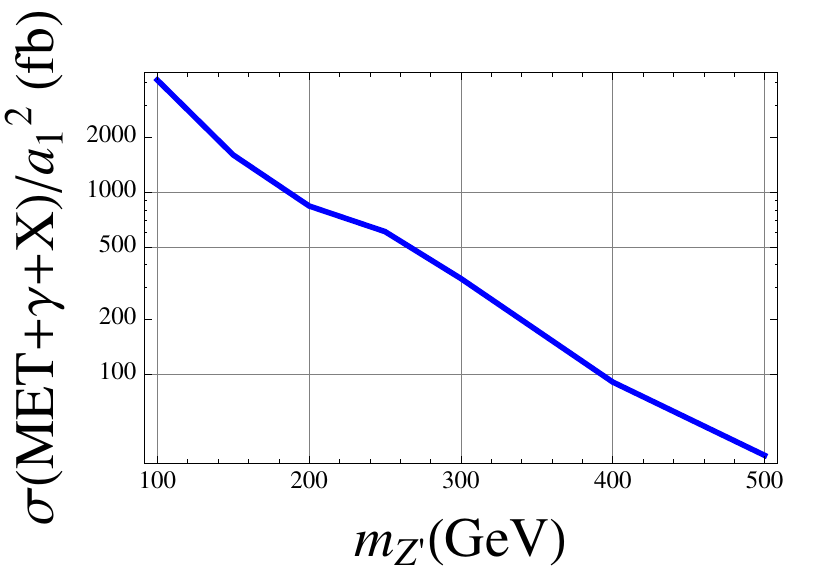}  
\caption{(Left) Acceptance for $pp\rightarrow \gamma \chi{\bar\chi}$ and other channels involving monophoton $+$ MET due to neutrinos as a function of $m_{Z'}$. (Right) Total cross section, $\sigma \times BR \times Acc$ (in fb), normalized to $a_1=1$.  
 }\label{exclusion8}
\end{figure}

In the first production mechanism, $p_T^{\gamma} = \slashed{E}_T$, up to radiation. Since $m_{DM}\simeq$ 150 GeV, one would expect large missing energy, and high efficiency on the cuts. We have obtained the efficiencies for the ATLAS and CMS cuts for the mass range $m_{Z'} \in [300,500]$ GeV, and those are in the 50-65\% range. Hence the monophoton searches are ideal to search for $Z'$ in this mass range. The blue(upper) line on the left of Fig.~\ref{exclusion8} shows the acceptance for the $pp\rightarrow \gamma\,\chi{\bar\chi}$ channel as a function of the $Z'$ mass at the LHC 8 TeV.

In the second production mechanism, one relies on the invisible branching ratio of the $Z$, which is 20 \%, and the cut on missing energy and high-$p_T$ photon may not be that efficient. Nevertheless, for events such as $p p \to Z' \gamma$ and $Z' \to \gamma Z$ the efficiency of the photon cut is order one, whereas the missing energy cut efficiency is reduced, as only one of the three bosons decay to neutrinos. The total efficiency is in the range 30-50\%, plus the reduction due to the branching ratio $Z\to \nu\bar{\nu}$.  Various lines on the left of Fig.~\ref{exclusion8} shows the acceptances for the individual channels giving rise to $pp\rightarrow X+\gamma+{\rm MET}$ as a function of the $Z'$ mass at the LHC 8 TeV. 

\begin{figure}[t]
\centering%
\includegraphics[scale=.8,natwidth=610,natheight=642]{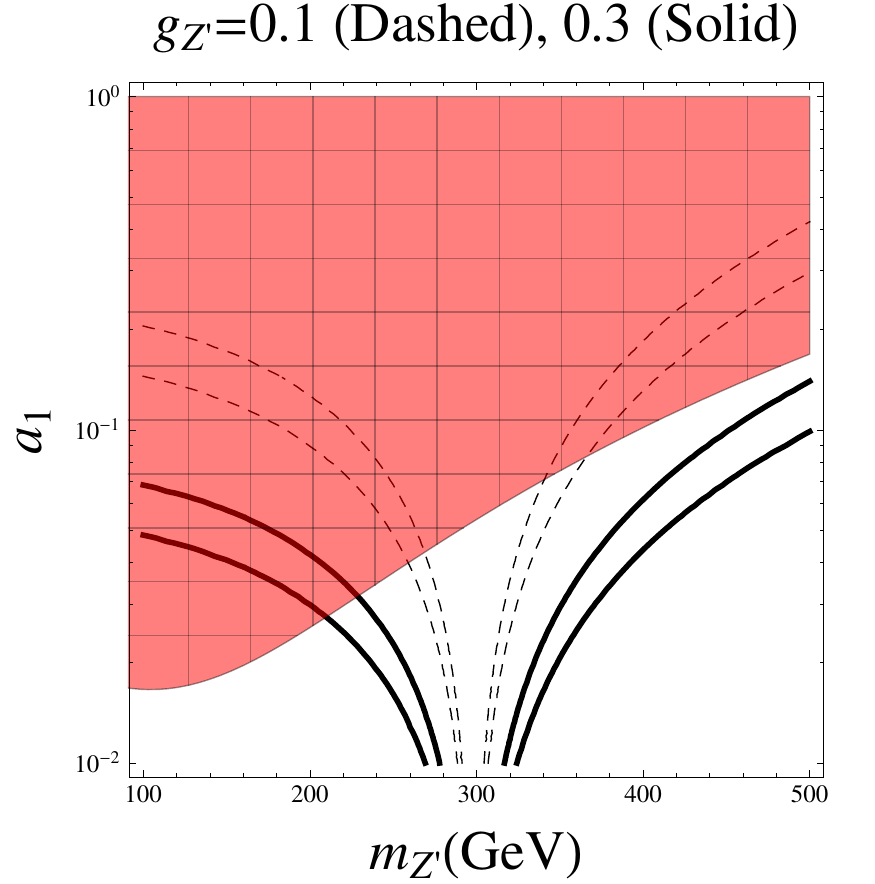}
\includegraphics[scale=.8,natwidth=610,natheight=642]{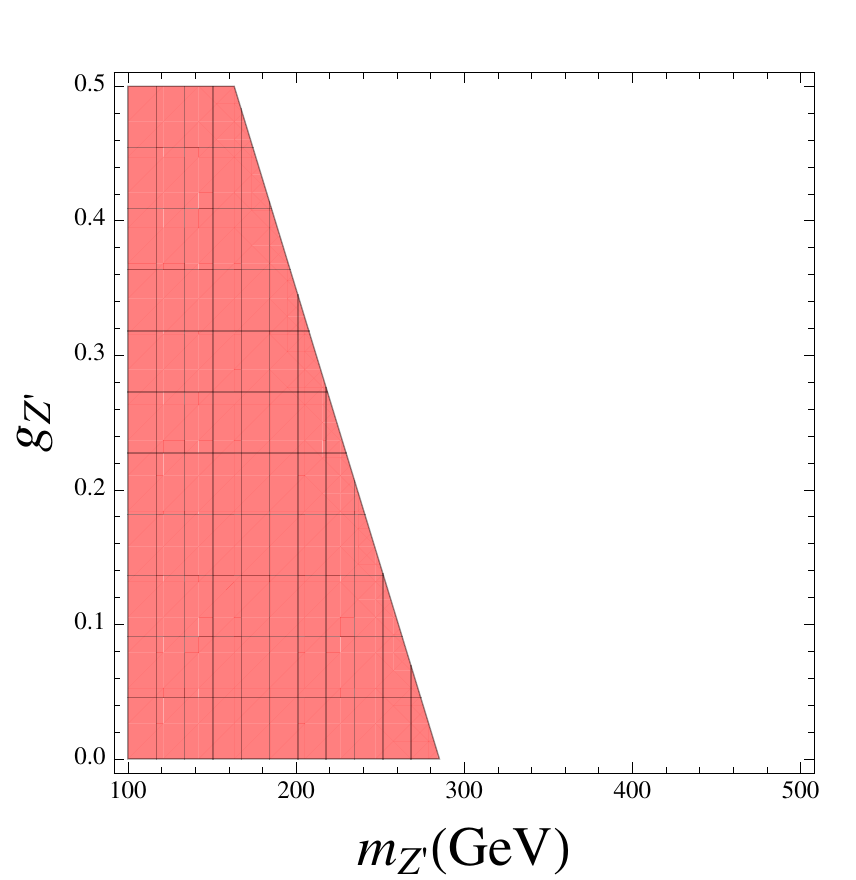} 
\caption{(Left)  Exclusion on the $(m_{Z'},a_1)$ parameter space with the overlay of the region satisfying $\langle\sigma v\rangle_{Z\gamma}/\langle \sigma v\rangle_0=0.04-0.08$ where $\langle\sigma v\rangle_0=3\times 10^{-26}\,{\rm cm}^3/{\rm s}$ is the thermal cross section. (Right)  Exclusion on the $(g_{Z'},m_{Z'})$ parameter space. In both plots, red region is excluded. 
}\label{exclusioncomb}
\end{figure}

To obtain the efficiencies we have 
 created new models in {\tt Feynrules}~\cite{Feynrules}. We then interfaced with  {\tt MadGraph}~\cite{MG5} using the UFO model format~\cite{UFO}.
We incorporated hadronization and showering effects using {\tt PYTHIA}~\cite{PYTHIA} and detector effects with 
 {\tt Delphes}~\cite{Delphes}. Finally, we have implemented the ATLAS and CMS cuts within {\tt ROOT}~\cite{ROOT}. In our simulation, jets are always anti-k$_T$ jets of size $R=0.5$.

From the monophoton searches at the LHC 8 TeV run, on the right of Fig.~\ref{exclusion8}, we apply the acceptance for $\gamma+{\rm MET}$ channels and show the total cross section ($\sigma \times BR \times Acc$) in fb, with re-scaled $a_1$. As a consequence, on the left of Fig.~\ref{exclusioncomb},  we present the exclusion limit on the $m_{Z'}$ vs $a_1$ plane by assuming that ${\rm Br} (Z'\rightarrow \chi {\bar \chi})=1$ above the mass threshold of a DM pair and overlay the region satisfying $\langle\sigma v\rangle_{Z\gamma}/\langle \sigma v\rangle_0=0.08-0.16$ where $\langle\sigma v\rangle_0=3\times 10^{-26}\,{\rm cm}^3/{\rm s}$ is the thermal cross section, for two values, $g_{Z'}=0.1, 0.3$. We find that the region away from the resonance in Fig.~\ref{fig:Zprime} can be excluded depending on the $Z'$ gauge coupling $g_{Z'}$: the excluded mass range is $m_{Z'}\lesssim 200\,{\rm GeV}$ for $g_{Z'}=0.3$, and $m_{Z'}\lesssim 260\,{\rm GeV}$ and $m_{Z'}\gtrsim 360\,{\rm GeV}$ for $g_{Z'}=0.1$.
On the right of Fig.~\ref{exclusioncomb},  we also show the excluded region in the $(m_{Z'}, g_{Z'})$ parameter space being compatible with the Fermi gamma-ray line.

We remark how the exclusion limit on the cross section changes when the extra decay mode of the $Z'$-boson such as $WW$ exists.
As discussed in the previous sections, the relic density condition requires extra DM annihilation channels than the one in $ZZ, Z\gamma$. 
In the low $Z'$ mass region below the threshold,  we can use the DM annihilation into a $Z'Z'$ pair with the $Z'$ gauge coupling for the correct relic density for $m_{Z'} > 2m_\chi - m_Z$ and the DM annihilation into $Z Z'$ channel with the extra CS coupling $a_2$ for $m_\chi < m_{Z'}< 2m_\chi -m_Z$.  These extra channels do not change the $Z'$ decay rate. 
However, in the high $Z'$ mass region above the threshold, we need extra channel, for instance, the $WW$ channel.
In this case, from eqs.~(\ref{ZpWW}) and (\ref{Zpchichi}), there is an extra contribution to the $Z'$ decay rate, which can be written as compared to the other dominant channels above the threshold as follows,
\bea
r\equiv \frac{\Gamma_{Z'}(WW)}{\Gamma_{Z'}(\chi{\bar\chi}) }=\bigg(\frac{b_1}{g_{Z'}}\bigg)^2 \frac{m^4_{Z'}}{m^2_W (m^2_{Z'}+2 m^2_\chi)}\,\frac{(1-4m^2_W/m^2_{Z'})^{5/2}}{(1-4m^2_\chi/m^2_{Z'})^{1/2}}.
\eea
As a result, when $b_1\sim g_{Z'}$, the $WW$ decay mode can be comparable to the decay mode into a DM pair, reducing the production cross section for $pp\rightarrow \gamma \,\chi{\bar\chi}$ above the threshold. But, the final states of the $WW$ decay can contain missing energy too and the efficiency of the corresponding channel is given by $Acc(WW)=28-47\%$ on the left of Fig.~\ref{exclusion8}, as compared to $Acc(\chi{\bar\chi})=50-65\%$ for the $pp\rightarrow \gamma\,\chi{\bar\chi}$ process. Then, the effective production cross section for $pp\rightarrow \gamma+{\rm MET}+X$ above the threshold is given by
\be
\sigma_{\rm eff}(pp\rightarrow \gamma+{\rm MET}+X)= \sigma(pp\rightarrow \gamma Z')\,\cdot \frac{Acc(\chi{\bar\chi})+r \,Acc(WW)}{1+r}.
\ee
Therefore, even for $r\sim 1$, the ``effective'' acceptance is given by $\frac{1}{2}(Acc(\chi{\bar\chi})+Acc(WW))=39-56\%$. Consequently, we still have a high efficiency for the wide range of the parameter space where the DM production cross section is reduced.

\subsection{Dedicated searches at 14 TeV}\label{14tev}

\begin{figure}[t]
\centering%
\includegraphics[width=7cm,natwidth=610,natheight=642]{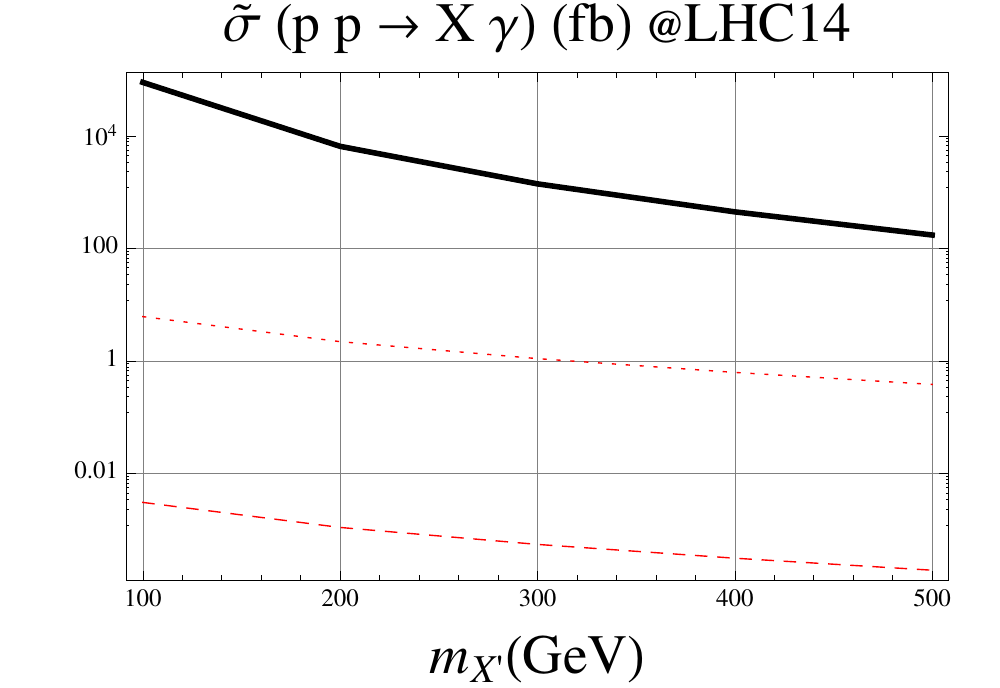}
\caption{Parton-level production cross section for $q{\bar q}\rightarrow (a, Z')+ \gamma$ at LHC $14\,{\rm TeV}$. Red-dashed line corresponds to the axion case, black-solid is the $Z'$ case. For illustration, we show the axion red-dotted line with $c_{1,2}=10$ and $v_s=55$ GeV.}
\label{assoprod14}
\end{figure}

When the LHC is up again in 2015, it may run at 13.5 TeV, and collect a luminosity per year of about 100 fb$^{-1}$. With the increase of energy, the cross section in associated production increases by a factor ${\cal O}(3-10)$, see Fig.~\ref{assoprod14}. Therefore, one could estimate that the sensitivity in signal could increase by a factor ${\cal O}$(10) per year, as compared to the 2012 run we discussed in the previous section. Keeping in mind that this is an optimistic estimate, it would mean that the reach in $a_1$ would increase by about factor three, and larger masses will be kinematically accessible.
 
In the low mass region ($m_X < 2 m_{\chi}$) a dedicated search would be more suitable. In this region, the efficiency to the missing energy cut in the ATLAS and CMS search is relatively low, and a bump search in the channel
\begin{equation}
 p \, p \to V^{*} \to V X \, \textrm{ with } X \to \gamma \gamma \ , 
\end{equation}
would be a more efficient way to look for the mediator $X=a,Z'$. One could also use the vector boson fusion channel, where the background rejection gain by cuts on $\Delta y_{j j}$ and hadronic activity could compensate the low parton luminosities of vector bosons fusing into the mediator.

Finally, let us mention that other channels, such as mono-Z or mono-lepton, could be used in association with the monophoton production we have discussed here. For a discussion, see Ref.~\cite{mono-Z}. 

\section{Conclusions}

We have considered the bounds of monophoton searches on the models of  the Fermi gamma-ray lines, focusing on the singlet fermion dark matter with a pseudo-scalar or vector mediator. Due to the suppression of the direct couplings between dark matter and the SM light quarks in the models,  the mediator coupling between dark matter and a pair of electroweak gauge bosons gives rise to a dominant production mechanism for the mediator field in association with a monophoton at the LHC. Thus, depending on the branching fraction of the invisible decay modes of the mediator, we have estimated the limit on the mediator production cross section with the LHC 8 TeV run and discussed the implications on the models.

In axion mediation,  where a sizable DM annihilation into a photon pair is obtained by the axion resonance, the LHC 8 TeV monophoton searches can rule out a certain region of the parameter space below $m_a=170\,{\rm GeV}$. Moreover, the fact that the axion mass is bounded to about $500\,{\rm GeV}$ within the resonance band could give a helpful guide to the monophoton searches at the LHC 14 TeV run.  On the other hand, in $Z'$ mediation, the DM annihilation into a single photon can be enhanced without a resonance, so a sizeable mediator coupling is allowed. In this case, the monophoton searches at the LHC 8 TeV already rules out some of the parameter space in the mediator mass and coupling.
 The dedicated searches for the $Z'$ or axion resonant production at the LHC 14 TeV are necessary and complementary to the monophoton searches discussed in this paper. In particular, a bump search on the $\gamma\gamma$ channel is especially promising in the low mass region.

\section*{Acknowledgments}
V. ~Sanz thanks the CERN TH group for its warm hospitality. 
M.~Park thanks for Olivier Mattelaer and Rakhi Mahbubani's help for the Monte Carlo simulations.
M.~Park is supported by the CERN-Korea fellowship through National
Research Foundation of Korea.

\def\theequation{A.\arabic{equation}}

\setcounter{equation}{0}

\vskip0.8cm
\noindent
{\Large \bf Appendix A: DM annihilation cross sections and $Z'$ decay rates in $Z'$ mediation} 
\vskip0.4cm
\noindent

In this appendix, we present the details of the computation of the DM annihilation cross sections and the $Z'$-boson decay rates in $Z'$ mediation.

From the Chern-Simons term (\ref{CSterm}), we compute the dark matter annihilation cross section into $Z\gamma$. The Feynman rules for the $Z'_{\mu}-Z_\nu-A_\sigma(p_2)$ vertex and $\chi-{\bar\chi}-Z'_\mu$ are $2a_1 \cos\theta_W \epsilon_{\mu\nu\rho\sigma}p^\rho_2$ and $i\,g_{Z'} \gamma_\mu/2$, respectively. Then, the scattering amplitude for $\chi(k_2){\bar\chi}(k_1)\rightarrow Z(p_1)\gamma(p_2)$ are
\be
{\cal M}_{\chi{\bar\chi}\rightarrow Z\gamma}=({\cal M}^{\lambda}_{\chi{\bar\chi}\rightarrow Z'})\,\bigg(\frac{i g_{\lambda\mu}}{s-m^2_{Z'}-i\Gamma_{Z'}m_{Z'}}\bigg)({\cal M}^\mu_{Z'\rightarrow Z\gamma}). \label{decayampl}
\ee
where
\bea
{\cal M}^{\lambda}_{\chi{\bar\chi}\rightarrow Z'}&=&\frac{i}{2}\,g_{Z'} {\bar v}(k_1)\gamma^\lambda u(k_2), \\
{\cal M}^\mu_{Z'\rightarrow Z\gamma}&=& 2 a_1\cos\theta_W\, \epsilon_{\mu\nu\rho\sigma} \epsilon^{*\nu}_Z(p_1) \epsilon^{*\sigma}_A(p_2) p^\rho_2.
\eea

Then, the annihilation cross section is given by
\be
\langle \sigma v\rangle_{Z\gamma}=\frac{1}{8\pi s} \Big|{\cal M}_{\chi{\bar\chi}\rightarrow Z\gamma}\Big|^2.  \label{annxsection}
\ee

The amplitude squared averaged over the spins of the dark matter is
\bea
\overline{|{\cal M}|^2}&=&\frac{g^2_{Z'}}{4} a^2_1 \cos^2\theta_W\, {\rm Tr}[( \slashed{k}_2+m_\chi)\gamma_{\lambda'}( \slashed{k}_1-m_\chi)\gamma_\lambda]\, \epsilon_{\mu\nu\rho\sigma} \epsilon_{\mu'\nu'\rho'\sigma'} g^{\lambda'\mu'} g^{\lambda\mu} \times\nonumber \\ 
&&\quad \times\sum_Z \epsilon^{*\nu}_Z(p_1)\epsilon^{\nu'}_Z(p_1)\, \sum_A \epsilon^{*\sigma}_A(p_2)\epsilon^{\sigma'}_A(p_2)\,p^\rho_2 \,p^{\rho'}_2\cdot \frac{1}{(s-m^2_{Z'})^2+\Gamma^2_{Z'}m^2_{Z'}}.
\eea
The trace part is calculated as
\be
{\rm Tr}[( \slashed{k}_2+m_\chi)\gamma^{\mu'}( \slashed{k}_1-m_\chi)\gamma^\mu]=4(k^\mu_1 k^{\mu'}_2+ k^\mu_2 k^{\mu'}_1)-4 (k_1\cdot k_2 +m^2_\chi) g^{\mu \mu'}.
\ee
We note that the polarization sums are
\bea
\sum_Z \epsilon^{*\mu}_Z(p)\epsilon^{\mu'}_Z(p)&=& -g^{\mu\nu'}+\frac{p^\mu p^{\mu'}}{m^2_Z}, \\
\sum_A \epsilon^{*\mu}_A(p)\epsilon^{\mu'}_A(p)&=& -g^{\mu\nu'}.
\eea
We also use the formulas for the contracted epsilon tensors as follows,
\bea
\epsilon_{\mu\nu\rho\sigma} \epsilon_{\mu'\nu'}\,^{\rho'\sigma'}g^{\mu \mu'}g^{\nu \nu'} g^{\rho \rho'}&=& -6 \delta^{\sigma'}_\sigma,  \label{eps1}\\
 \epsilon_{\mu\nu\rho\sigma} \epsilon_{\mu'\nu'}\,^{\rho'\sigma'}g^{\mu \mu'}g^{\nu \nu'} &=& -2 (\delta^{\rho'}_{\rho} \delta^{\sigma'}_\sigma -\delta^{\sigma'}_{\rho} \delta^{\rho'}_\sigma),  \label{eps2}\\
  \epsilon_{\mu\nu\rho\sigma} \epsilon_{\mu'\nu'}\,^{\rho'\sigma'}g^{\mu \mu'}&=& -\delta_\sigma^{\sigma'} \delta_\nu^{\nu'} \delta_\rho^{\rho'}+\delta_\sigma^{\nu'}\delta_\nu^{\sigma'}\delta_\rho^{\rho'}+\delta_\sigma^{\sigma'} \delta_\nu^{\rho'} \delta_\rho^{\nu'} \nonumber \\
  &&-\delta_\sigma^{\rho'}\delta_\nu^{\sigma'}\delta_\rho^{\nu'}-\delta_\sigma^{\nu'} \delta_\nu^{\rho'} \delta_\rho^{\sigma'} +\delta_\sigma^{\rho'} \delta_\nu^{\nu'} \delta_\rho^{\sigma'}.  \label{eps3}
\eea
Using the above formulas, we get
\be
 \epsilon_{\mu\nu\rho\sigma} \epsilon_{\mu'\nu'\rho'\sigma'} g^{\mu \mu'} \Big(g^{\nu \nu'} -\frac{p^\nu_1 p^{\nu'}_1}{m^2_Z}\Big) g^{\sigma\sigma'} p_2^\rho p_2^{\rho'}
=-6 p^2_2 +\frac{2}{m^2_Z} \Big(p^2_1 p^2_2-(p_1\cdot p_2)^2\Big),
\ee
$$
\epsilon_{\mu\nu\rho\sigma} \epsilon_{\mu'\nu'\rho'\sigma'}(k^\mu_1 k^{\mu'}_2+ k^\mu_2 k^{\mu'}_1)\Big(g^{\nu \nu'}-\frac{p^\nu_1 p^{\nu'}_1}{m^2_Z}\Big)g^{\sigma\sigma'} p^\rho_2 p^{\rho'}_2
$$
\bea
&=&-4\Big((k_1\cdot k_2)p^2_2 - (k_1\cdot p_2)(k_2\cdot p_2)\Big)-\frac{2}{m^2_Z}\Big(-(k_1\cdot k_2) p^2_1 p^2_2 +(k_1\cdot p_1)(k_2\cdot p_1) p^2_2 \nonumber \\
&&\quad+(k_1\cdot k_2) (p_1\cdot p_2)^2
-(k_1\cdot p_2)(k_2\cdot p_1)(p_1\cdot p_2)\nonumber \\
&&\quad -(k_1\cdot p_1)(k_2\cdot p_2)(p_1\cdot p_2)
+(k_1\cdot p_2)(k_2\cdot p_2) p^2_1\Big).
\eea
Therefore, for $p^2_2=0$, we get 
\bea
\overline{|{\cal M}|^2}&=&\frac{g^2_{Z'}}{4} a^2_1 \cos^2\theta_W\cdot
\frac{1}{(s-m^2_{Z'})^2+\Gamma^2_{Z'}m^2_{Z'}}\times \nonumber \\
&&\times \bigg[ \frac{s}{m^2_Z}(s-m^2_Z)^2+16(k_1\cdot p_2)(k_2\cdot p_2)\nonumber \\
&&\quad -\frac{8}{m^2_Z}\Big(k_1 (p_1\cdot p_2)-p_1(k_1\cdot p_2)\Big)\Big(k_2 (p_1\cdot p_2)-p_1(k_2\cdot p_2)\Big)\bigg].
\eea
For non-relativistic dark matter, we get $s\simeq 4 m^2_\chi$, so the squared amplitude becomes
\bea
\overline{|{\cal M}|^2}\simeq \frac{g^2_{Z'}}{4}\frac{a^2_1 \cos^2\theta_W}{(4m^2_\chi-m^2_Z)^2+\Gamma^2_{Z'}m^2_{Z'}}\, (4m^2_\chi-m^2_Z)^2\Big(\frac{4m^2_{\chi}}{m^2_Z}+1\Big).
\eea
Then, including the phase factor by $\Big|{\cal M}_{\chi{\bar\chi}\rightarrow Z\gamma}\Big|^2=(1-m^2_Z/s)\,\overline{|{\cal M}|^2}$ and using eq.~(\ref{annxsection}), we get the annihilation cross section as
\be
\langle\sigma v\rangle_{Z\gamma}=\frac{g^2_{Z'} a^2_1 \cos^2\theta_W}{32\pi} \frac{16 m^4_\chi}{(4m^2_\chi-m^2_Z)^2+\Gamma^2_{Z'}m^2_{Z'}}\,\left(1-\frac{m^2_Z}{4m^2_\chi}\right)^3 \Big(\frac{1}{m^2_Z}+\frac{1}{4m^2_\chi}\Big).
\ee
In a similar matter, as given in the text, we can obtain the annihilation cross sections for $\chi{\bar\chi}\rightarrow ZZ, ZZ', WW$ from the CS terms in eq.~(\ref{CSterm}) and $\chi{\bar\chi}\rightarrow h\gamma$ from the CS coupling for the Higgs in eq.~(\ref{CSHiggs}). Furthermore, when $m_{Z'}<m_\chi$, there is a t-channel annihilation channel for $\chi{\bar \chi}\rightarrow Z' Z'$ due to the $Z'$ gauge coupling to dark matter.

Now we consider the decay rate of the $Z'$ gauge boson.
The decay rate is given in terms of the decay amplitude, ${\cal M}_{Z'\rightarrow Z\gamma}=\epsilon^\mu_{Z'}(p_1+p_2){\cal M}^\mu_{Z'\rightarrow Z\gamma}$, by
 \be
 \Gamma_{Z'}(Z'\rightarrow Z\gamma)=\frac{1}{16\pi m_{Z'}}\, \bigg|{\cal M}_{Z'\rightarrow Z\gamma}\bigg|^2. \label{decayrate}
 \ee
The squared decay amplitude averaged over the spins of the $Z'$ gauge boson is
\bea
\overline{|{\cal M}|^2}&=&\frac{4}{3} a^2_1 \cos^2\theta_W\,\epsilon_{\mu\nu\rho\sigma} \epsilon_{\mu'\nu'\rho'\sigma'}\sum_{Z'} \epsilon^{*\mu'}_{Z'}(p_1+p_2)\epsilon^{\mu}_{Z'}(p_1+p_2)\times \nonumber \\
&&\quad \times \sum_Z \epsilon^{*\nu}_Z(p_1)\epsilon^{\nu'}_Z(p_1)\, \sum_A \epsilon^{*\sigma}_A(p_2)\epsilon^{\sigma'}_A(p_2)\,p^\rho_2 \,p^{\rho'}_2.
\eea
Then, using eqs.~(\ref{eps1})-(\ref{eps3}) and $p^2_2=0$, we get 
\be
\overline{|{\cal M}|^2}=\frac{2}{3} a^2_1 \cos^2\theta_W\, (m^2_{Z'}-m^2_Z)^2\Big(\frac{1}{m^2_Z}+\frac{1}{m^2_{Z'}}\Big).
\ee
Therefore, including the phase factor by $\Big|{\cal M}_{Z'\rightarrow Z\gamma}\Big|^2=(1-m^2_Z/m^2_{Z'})\,\overline{|{\cal M}|^2}$ and using eq.~(\ref{decayrate}), we get the decay rate as
\be
\Gamma_{Z'}(Z'\rightarrow Z\gamma)=\frac{m^3_{Z'}}{24\pi}\, a^2_1 \cos^2\theta_W\,\left(1-\frac{m^2_Z}{m^2_{Z'}}\right)^3 \Big(\frac{1}{m^2_Z}+\frac{1}{m^2_{Z'}}\Big).
\ee
Similarly, as given in the text, we can get the decay rates for $Z'\rightarrow ZZ, WW, \chi {\bar\chi}$  from the CS coupling in eq.~(\ref{CSterm}) and for $Z'\rightarrow h\gamma, hZ, t{\bar t}$ from the CS term for the Higgs 
in eq.~(\ref{CSHiggs}) and top quark coupling to $Z'$.

\def\theequation{B.\arabic{equation}}

\setcounter{equation}{0}

\vskip0.8cm
\noindent
{\Large \bf Appendix B: Mediator production cross sections at the LHC}\label{appZp}
\vskip0.4cm
\noindent

In this appendix, we present the details on the derivation of the production cross sections for the axion and $Z'$ gauge boson in association with a monophoton at the LHC.

First, in axion mediation, we compute the Drell-Yan production cross section for the axion in association with electroweak gauge boson at the LHC: $q{\bar q}'\rightarrow V^*_1\rightarrow a\,V_2$. For the quark gauge coupling, $g_f V_{1,\mu} {\bar f}'\gamma^\mu (v_f-a_f \gamma^5)f$, and the axion coupling, 
$c_{V_1 V_2} \epsilon_{\mu\nu\rho\sigma} a\, F^{\mu\nu}_{V_1}F^{\rho\sigma}_{V_2}$, ignoring the quark masses, we get the squared scattering amplitude for $q(k_2){\bar q}'(k_1)\rightarrow V^*_1\rightarrow a(p_1)\,V_2(p_2)$ as
\bea
\overline{|{\cal M}|^2}&=&\frac{1}{N_c}\, \frac{4s^2_V \,g^2_f c^2_{V_1 V_2}}{(s-m^2_{V_1})^2+m^2_{V_1}\Gamma^2_{V_1}}\, {\rm Tr}\Big[ \slashed{k}_2\gamma^{\nu'} \slashed{k}_1\gamma^\nu (v^2_f+a^2_f-2v_f a_f \gamma^5)\Big] \times \nonumber \\
&&\quad\times \epsilon_{\mu\nu\rho\sigma}\epsilon_{\mu'\nu'\rho'\sigma'}\sum_{V_2}\epsilon^{*\sigma}_{V_2}(p_2) \epsilon^{\sigma'}_{V_2}(p_2)\, p^\mu_3 p^{\mu'}_3 p^\rho_2 p^{\rho'}_2
\eea
where $s_V$ is the symmetric factor, which is equal to $1$ for $V_1\neq V_2$ and $2$ for $V_1=V_2$.
Thus, after calculating the trace and the Lorentz contracted quantities, we obtain
\bea
\overline{|{\cal M}|^2}&=& \frac{1}{N_c}\,\frac{32s^2_V \,g^2_f c^2_{V_1 V_2}(v^2_f+a^2_f)}{(s-m^2_{V_1})^2+m^2_{V_1}\Gamma^2_{V_1}}\, \bigg[ (p_1\cdot p_2)\bigg\{ (k_2\cdot p_1)(k_1\cdot p_2)+(k_2\cdot p_2)(k_1\cdot p_1) \bigg\} \nonumber \\
&&\quad -p^2_1 (k_1\cdot p_2)(k_2\cdot p_2)-p^2_2 (k_2\cdot p_1)(k_1\cdot p_1)\bigg].
\eea
Consequently, we get the production cross section as follows,
\bea
\sigma(q{\bar q}'\rightarrow a\,V_2)&=& \frac{1}{32\pi^2 s}\int d\Omega\,\frac{|{\vec p}|}{\sqrt{s}} \,\overline{|{\cal M}|^2} \nonumber \\
&=&\frac{1}{N_c \pi}\,\frac{s^2_V\,g^2_f c^2_{V_1 V_2}(v^2_f+a^2_f)}{(s-m^2_{V_1})^2+m^2_{V_1}\Gamma^2_{V_1}}\, \frac{|{\vec p}|}{\sqrt{s}}\bigg[(s-m^2_a-m^2_{V_2})\Big(E_1 E_2+\frac{1}{3}|{\vec p}|^2\Big)
\nonumber \\
&&\quad -m^2_a \Big(E^2_2-\frac{1}{3}|{\vec p}|^2\Big)-m^2_{V_2}\Big(E^2_1-\frac{1}{3}|{\vec p}|^2\Big)\bigg]
\eea
where
\bea
|{\vec p}|&\equiv &\frac{\sqrt{s}}{2}\Big(1-\frac{(m_a+m_{V_2})^2}{s}\Big)^{1/2}\Big(1-\frac{(m_a-m_{V_2})^2}{s}\Big)^{1/2},\\
E_1&\equiv & \frac{\sqrt{s}}{2}\Big(1+\frac{m^2_a-m^2_{V_2}}{s}\Big),\\
E_2&\equiv & \frac{\sqrt{s}}{2}\Big(1-\frac{m^2_a-m^2_{V_2}}{s}\Big).
\eea

In particular,  for $q{\bar q}\rightarrow \gamma^*\rightarrow a\,\gamma$, we get $s_V=2$, so the production cross section for the process is
\bea
\sigma(q{\bar q}\rightarrow a\,\gamma)&=& \frac{8Q^2_q }{3N_c}\,\frac{(c_1+c_2)^2\alpha^3}{(16\pi v_s)^2}\Big(1-\frac{m^2_a}{s}\Big)^3
\eea
with $Q_q$ being the electromagnetic charge of the quark.
If the axion decays into a DM pair, there are a single photon plus missing energy in the final states. 
The full cross section for $q{\bar q}\rightarrow a\,\gamma$ is 
\bea
\sigma(q{\bar q}\rightarrow a\,\gamma)&=& \frac{2\alpha}{3N_c}\bigg[ 4 Q^2_q c^2_{\gamma\gamma}+\frac{(v^2_q+a^2_q) c^2_{Z\gamma}}{\sin^2(2\theta_W)}\, \frac{s^2}{(s-m^2_Z)^2+\Gamma^2_Z m^2_Z} \nonumber \\
&&\quad+ \frac{4Q_q v_q c_{\gamma\gamma} c_{Z\gamma}}{\sin(2\theta_W) }\,\frac{s(s-m^2_Z)}{(s-m^2_Z)^2+\Gamma^2_Z m^2_Z} \bigg] \Big(1-\frac{m^2_a}{s}\Big)^3
\eea
where $v_q=T^q_3(1-4|Q_q|\sin^2\theta_W)$ and $a_q=T^q_3$, that is, 
\be
v_u=\frac{1}{2}\Big(1-\frac{8}{3}\sin^2\theta_W\Big), \,\, v_d=\frac{1}{2}\Big(-1+\frac{4}{3}\sin^2\theta_W\Big), \,\, a_u=-a_d=\frac{1}{2}.
\ee

Second, in $Z'$ mediation, we compute the Drell-Yan production cross section for $Z' (Z)$ in association with a monophoton at the LHC: $q{\bar q}\rightarrow V^*_3\rightarrow  V_1\,V_2$. For the quark gauge coupling, $g_q V_{3,\mu} {\bar q}\gamma^\mu (v_q-a_q \gamma^5)q$, and the CS coupling, 
$a_1 \epsilon^{\mu\nu\rho\sigma} Z'_\mu  Z_\nu F^Y_{\rho\sigma}$, ignoring the quark masses, we get the squared scattering amplitude for $q(k_2){\bar q}(k_1)\rightarrow V^*_3\rightarrow V_1(p_1)\,V_2(p_2)$ as
\bea
\overline{|{\cal M}|^2}&=&\frac{1}{N_c}\, \frac{g^2_q a^2_1 \cos^2\theta_W}{(s-m^2_{V_3})^2+m^2_{V_3}\Gamma^2_{V_3}}\, {\rm Tr}\Big[ \slashed{k}_2\gamma^{\nu'} \slashed{k}_1\gamma^\nu (v^2_q+a^2_q-2v_q a_q \gamma^5)\Big] \times \nonumber \\
&&\quad\times \epsilon_{\mu\nu\rho\sigma}\epsilon_{\mu'\nu'\rho'\sigma'}\sum_{V_2}\epsilon^{*\sigma}_{V_2}(p_2) \epsilon^{\sigma'}_{V_2}(p_2)\, \sum_{V_1}\epsilon^{*\sigma}_{V_1}(p_1) \epsilon^{\sigma'}_{V_1}(p_1)\, p^\rho_2 p^{\rho'}_2.
\eea
Thus, after calculating the trace and the Lorentz contracted quantities, we obtain
\bea
\overline{|{\cal M}|^2}&=& \frac{1}{N_c}\,\frac{8g^2_q a^2_1\cos^2\theta_W (v^2_q+a^2_q)}{(s-m^2_{V_3})^2+m^2_{V_3}\Gamma^2_{V_3}}\, \bigg[ 
p^2_2 (k_1\cdot k_2)+2(k_1\cdot p_2)(k_2\cdot p_2) \nonumber \\
&&\quad+ \frac{1}{m^2_{V_1}}\bigg\{(p_1\cdot p_2)\Big( (k_2\cdot p_1)(k_1\cdot p_2)+(k_2\cdot p_2)(k_1\cdot p_1) \Big) \nonumber \\
&&\quad\quad -p^2_1 (k_1\cdot p_2)(k_2\cdot p_2)-p^2_2 (k_2\cdot p_1)(k_1\cdot p_1)\bigg\}   \bigg].
\eea
Consequently,  we get the production cross section as follows,
\bea
\sigma(q{\bar q}\rightarrow Z'\,\gamma)&=&\frac{1}{24\pi N_c}\frac{g^2_q a^2_1\cos^2\theta_W  (v^2_q+a^2_q)}{(s-m^2_Z)^2+m^2_Z\Gamma^2_Z}\, s  \bigg( 1+\frac{s}{m^2_{Z'}}\bigg)\bigg(1-\frac{m^2_{Z'}}{s}  \bigg)^3. 
\eea
with $g_q=\frac{g}{2\cos\theta_W}$.

When the SM light quarks couple to $Z'$ boson through a kinetic mixing $\epsilon$  between $Z'$ boson and Z-boson, we get $g_q=\epsilon g_{Z'}$ and $v_q=1$ and $a_q=0$.
Then, we also obtain the Drell-Yan production cross section for $Z$ in association with a monophoton as
\bea
\sigma(q{\bar q}\rightarrow Z\,\gamma)=\frac{1}{24\pi N_c}\frac{ \epsilon^2 g^2_{Z'}a^2_1\cos^2\theta_W }{(s-m^2_{Z'})^2+m^2_{Z'}\Gamma^2_{Z'}}\, s  \bigg( 1+\frac{s}{m^2_{Z}}\bigg)\bigg(1-\frac{m^2_{Z}}{s}  \bigg)^3. 
\eea
In the text, we consider this process to be subdominant, as compared to the $Z'\gamma$ channel.


\begin{thebibliography}{999}

  
\bibitem{fermi}
%\cite{Atwood:2009ez}
%\bibitem{Atwood:2009ez}
  W.~B.~Atwood {\it et al.}  [LAT Collaboration],
  %``The Large Area Telescope on the Fermi Gamma-ray Space Telescope Mission,''
  Astrophys.\ J.\  {\bf 697} (2009) 1071
  [arXiv:0902.1089 [astro-ph.IM]];
  %%CITATION = ARXIV:0902.1089;%%
%\cite{Abdo:2010nc}
%\bibitem{Abdo:2010nc}
  A.~A.~Abdo {\it et al.},
  %``Fermi LAT Search for Photon Lines from 30 to 200 GeV and Dark Matter
  %Implications,''
  Phys.\ Rev.\ Lett.\  {\bf 104} (2010) 091302
  [arXiv:1001.4836 [astro-ph.HE]];
  %%CITATION = PRLTA,104,091302;%%
%\cite{Ackermann:2012qk}
%\bibitem{Ackermann:2012qk}
  F.~M.~Ackermann {\it et al.}  [LAT Collaboration],
  %``Fermi LAT Search for Dark Matter in Gamma-ray Lines and the Inclusive Photon Spectrum,''
  arXiv:1205.2739 [astro-ph.HE].
  %%CITATION = ARXIV:1205.2739;%%


\bibitem{dwarfgalaxy}
%\cite{Ackermann:2011wa}
%\bibitem{Ackermann:2011wa}
  M.~Ackermann {\it et al.}  [Fermi-LAT Collaboration],
  %``Constraining Dark Matter Models from a Combined Analysis of Milky Way Satellites with the Fermi Large Area Telescope,''
  Phys.\ Rev.\ Lett.\  {\bf 107} (2011) 241302
  [arXiv:1108.3546 [astro-ph.HE]].
  %%CITATION = ARXIV:1108.3546;%%



\bibitem{weniger}
%\cite{Weniger:2012tx}
%\bibitem{Weniger:2012tx}
  C.~Weniger,
  %``A Tentative Gamma-Ray Line from Dark Matter Annihilation at the Fermi Large Area Telescope,''
  JCAP {\bf 1208} (2012) 007
  [arXiv:1204.2797 [hep-ph]].
  %%CITATION = ARXIV:1204.2797;%%




\bibitem{otheranalysis}  
%\cite{Bringmann:2012vr}
%\bibitem{Bringmann:2012vr}
  T.~Bringmann, X.~Huang, A.~Ibarra, S.~Vogl and C.~Weniger,
  %``Fermi LAT Search for Internal Bremsstrahlung Signatures from Dark Matter Annihilation,''
  arXiv:1203.1312 [hep-ph]; 
  %%CITATION = ARXIV:1203.1312;%%
%\cite{Tempel:2012ey}
%\bibitem{Tempel:2012ey}
  E.~Tempel, A.~Hektor and M.~Raidal,
  %``Fermi 130 GeV gamma-ray excess and dark matter annihilation in sub-haloes and in the Galactic centre,''
  arXiv:1205.1045 [hep-ph];
  %%CITATION = ARXIV:1205.1045;%%
%\cite{Su:2012ft}
%\bibitem{Su:2012ft}
  M.~Su and D.~P.~Finkbeiner,
  %``Strong Evidence for Gamma-ray Line Emission from the Inner Galaxy,''
  arXiv:1206.1616 [astro-ph.HE].
  %%CITATION = ARXIV:1206.1616;%%


\bibitem{FermiSymposium}
A. Albert, Talk at the Fermi Symposium, November 2, 2012, California, USA.



\bibitem{earthlimb}
%\cite{Hektor:2012ev}
%\bibitem{Hektor:2012ev}
  A.~Hektor, M.~Raidal and E.~Tempel,
  %``Fermi-LAT gamma-ray signal from Earth Limb, systematic detector effects and their implications for the 130 GeV gamma-ray excess,''
  arXiv:1209.4548 [astro-ph.HE];
  %%CITATION = ARXIV:1209.4548;%%
%\cite{Finkbeiner:2012ez}
%\bibitem{Finkbeiner:2012ez}
  D.~P.~Finkbeiner, M.~Su and C.~Weniger,
  %``Is the 130 GeV Line Real? A Search for Systematics in the Fermi-LAT Data,''
  arXiv:1209.4562 [astro-ph.HE].
  %%CITATION = ARXIV:1209.4562;%%


\bibitem{xenon}
%\cite{Aprile:2012nq}
%\bibitem{Aprile:2012nq}
  E.~Aprile {\it et al.}  [XENON100 Collaboration],
  %``Dark Matter Results from 225 Live Days of XENON100 Data,''
  Phys.\ Rev.\ Lett.\  {\bf 109} (2012) 181301
  [arXiv:1207.5988 [astro-ph.CO]].
  %%CITATION = ARXIV:1207.5988;%%



\bibitem{scalarDM}
%\cite{Cline:2012nw}
%\bibitem{Cline:2012nw}
  J.~M.~Cline,
  %``130 GeV dark matter and the Fermi gamma-ray line,''
  arXiv:1205.2688 [hep-ph].
 


 \bibitem{axion1} 
%\cite{Lee:2012bq}
%\bibitem{Lee:2012bq}
  H.~M.~Lee, M.~Park and W.~-I.~Park,
  %``Fermi Gamma Ray Line at 130 GeV from Axion-Mediated Dark Matter,''
  Phys.\ Rev.\ D {\bf 86} (2012) 103502
  [arXiv:1205.4675 [hep-ph]].
  %%CITATION = ARXIV:1205.4675;%%



\bibitem{axion2}
%\cite{Lee:2012wz}
%\bibitem{Lee:2012wz}
  H.~M.~Lee, M.~Park and W.~-I.~Park,
  %``Axion-mediated dark matter and Higgs diphoton signal,''
  arXiv:1209.1955 [hep-ph]. 
  %%CITATION = ARXIV:1209.1955;%%




\bibitem{zprimeCS}
%\cite{Dudas:2012pb}
%\bibitem{Dudas:2012pb}
  E.~Dudas, Y.~Mambrini, S.~Pokorski and A.~Romagnoni,
  %``Extra U(1) as natural source of a monochromatic gamma ray line,''
  arXiv:1205.1520 [hep-ph];
  %%CITATION = ARXIV:1205.1520;%%
%\cite{Mambrini:2009ad}
%\bibitem{Mambrini:2009ad}
  Y.~Mambrini,
  %``A Clear Dark Matter gamma ray line generated by the Green-Schwarz mechanism,''
  JCAP {\bf 0912} (2009) 005
  [arXiv:0907.2918 [hep-ph]];
  %%CITATION = ARXIV:0907.2918;%%
%\cite{Dudas:2009uq}
%\bibitem{Dudas:2009uq}
  E.~Dudas, Y.~Mambrini, S.~Pokorski and A.~Romagnoni,
  %``(In)visible Z-prime and dark matter,''
  JHEP {\bf 0908} (2009) 014
  [arXiv:0904.1745 [hep-ph]].
  %%CITATION = ARXIV:0904.1745;%%


\bibitem{zprimeTquark}
%\cite{Jackson:2009kg}
%\bibitem{Jackson:2009kg}
  C.~B.~Jackson, G.~Servant, G.~Shaughnessy, T.~M.~P.~Tait and M.~Taoso,
  %``Higgs in Space!,''
  JCAP {\bf 1004} (2010) 004
  [arXiv:0912.0004 [hep-ph]].
  %%CITATION = ARXIV:0912.0004;%%


\bibitem{othermodels}
%\cite{Ibarra:2012dw}
%\bibitem{Ibarra:2012dw}
  A.~Ibarra, S.~Lopez Gehler and M.~Pato,
  %``Dark matter constraints from box-shaped gamma-ray features,''
  arXiv:1205.0007 [hep-ph];
  %%CITATION = ARXIV:1205.0007;%%
   %%CITATION = ARXIV:1205.2688;%%
%\cite{Choi:2012ap}
%\bibitem{Choi:2012ap}
  K.~-Y.~Choi and O.~Seto,
  %``A Dirac right-handed sneutrino dark matter and its signature in the gamma-ray lines,''
  arXiv:1205.3276 [hep-ph];
  %%CITATION = ARXIV:1205.3276;%%
%\cite{Kyae:2012vi}
%\bibitem{Kyae:2012vi}
  B.~Kyae and J.~-C.~Park,
  %``130 GeV Gamma-Ray Line from Dark Matter Decay,''
  arXiv:1205.4151 [hep-ph];
  %%CITATION = ARXIV:1205.4151;%%
  %\cite{Rajaraman:2012db}
%\bibitem{Rajaraman:2012db}
  A.~Rajaraman, T.~M.~P.~Tait and D.~Whiteson,
  %``Two Lines or Not Two Lines? That is the Question of Gamma Ray Spectra,''
  arXiv:1205.4723 [hep-ph];
  %%CITATION = ARXIV:1205.4723;%%
  %\cite{Acharya:2012dz}
%\bibitem{Acharya:2012dz}
  B.~S.~Acharya, G.~Kane, P.~Kumar, R.~Lu and B.~Zheng,
  %``Mixed Wino-Axion Dark Matter in String/M Theory and the 130 GeV Gamma-line 'Signal',''
  arXiv:1205.5789 [hep-ph];
  %%CITATION = ARXIV:1205.5789;%%
%\cite{Buckley:2012ws}
%\bibitem{Buckley:2012ws}
  M.~R.~Buckley and D.~Hooper,
  %``Implications of a 130 GeV Gamma-Ray Line for Dark Matter,''
  Phys.\ Rev.\ D {\bf 86} (2012) 043524
  [arXiv:1205.6811 [hep-ph]];
  %%CITATION = ARXIV:1205.6811;%%
  %\cite{Das:2012ys}
%\bibitem{Das:2012ys}
  D.~Das, U.~Ellwanger and P.~Mitropoulos,
  %``A 130 GeV photon line from dark matter annihilation in the NMSSM,''
  JCAP {\bf 1208} (2012) 003
  [arXiv:1206.2639 [hep-ph]];
  %%CITATION = ARXIV:1206.2639;%%
%\cite{Kang:2012bq}
%\bibitem{Kang:2012bq}
  Z.~Kang, T.~Li, J.~Li and Y.~Liu,
  %``Brightening the (130 GeV) Gamma-Ray Line,''
  arXiv:1206.2863 [hep-ph];
  %%CITATION = ARXIV:1206.2863;%%  
  %\cite{Park:2012xq}
%\bibitem{Park:2012xq}
  J.~-C.~Park and S.~C.~Park,
  %``Radiatively decaying scalar dark matter through U(1) mixings and the Fermi 130 GeV gamma-ray line,''
  arXiv:1207.4981 [hep-ph];
  %%CITATION = ARXIV:1207.4981;%%
%\cite{Tulin:2012uq}
%\bibitem{Tulin:2012uq}
  S.~Tulin, H.~-B.~Yu and K.~M.~Zurek,
  %``Three Exceptions for Thermal Dark Matter with Enhanced Annihilation to Gamma Gamma,''
  arXiv:1208.0009 [hep-ph];
  %%CITATION = ARXIV:1208.0009;%%
  %\cite{Li:2012jf}
%\bibitem{Li:2012jf}
  T.~Li, J.~A.~Maxin, D.~V.~Nanopoulos and J.~W.~Walker,
  %``A 125.5 GeV Higgs Boson in F-SU(5): Imminently Observable Proton Decay, A 130 GeV Gamma-ray Line, and SUSY Multijets & Light Stops at the LHC8,''
  arXiv:1208.1999 [hep-ph];
  %%CITATION = ARXIV:1208.1999;%%
%\cite{Cline:2012bz}
%\bibitem{Cline:2012bz}
  J.~M.~Cline, A.~R.~Frey and G.~D.~Moore,
  %``Composite magnetic dark matter and the 130 GeV line,''
  arXiv:1208.2685 [hep-ph];
    %%CITATION = ARXIV:1208.2685;%%
    %\cite{Bai:2012qy}
%\bibitem{Bai:2012qy}
  Y.~Bai and J.~Shelton,
  %``Gamma Lines without a Continuum: Thermal Models for the Fermi-LAT 130 GeV Gamma Line,''
  arXiv:1208.4100 [hep-ph];
  %%CITATION = ARXIV:1208.4100;%%
    %\cite{Bergstrom:2012fu}
%\bibitem{Bergstrom:2012fu}
  L.~Bergstrom,
  %``The 130 GeV Fingerprint of Right-Handed Neutrino Dark Matter,''
  arXiv:1208.6082 [hep-ph];
  %%CITATION = ARXIV:1208.6082;%%
  %\cite{Wang:2012uy}
%\bibitem{Wang:2012uy}
  L.~Wang and X.~-F.~Han,
  %``130 GeV gamma-ray line and enhancement of $h\to\gamma\gamma$ in the Higgs triplet model plus a scalar dark matter,''
  arXiv:1209.0376 [hep-ph]; 
  %%CITATION = ARXIV:1209.0376;%%
%\cite{SchmidtHoberg:2012ip}
%\bibitem{SchmidtHoberg:2012ip}
  K.~Schmidt-Hoberg, F.~Staub and M.~W.~Winkler,
  %``Enhanced diphoton rates at Fermi and the LHC,''
  arXiv:1211.2835 [hep-ph];
  %%CITATION = ARXIV:1211.2835;%%
  %\cite{Farzan:2012kk}
%\bibitem{Farzan:2012kk}
  Y.~Farzan and A.~R.~Akbarieh,
  %``Natural explanation for 130 GeV photon line within vector boson dark matter model,''
  arXiv:1211.4685 [hep-ph].
  %%CITATION = ARXIV:1211.4685;%%
%\cite{Chalons:2012xf}
%\bibitem{Chalons:2012xf}
  G.~Chalons, M.~J.~Dolan and C.~McCabe,
  %``Neutralino dark matter and the Fermi gamma-ray lines,''
  arXiv:1211.5154 [hep-ph].
  %%CITATION = ARXIV:1211.5154;%%



\bibitem{CMS-monoa}
 S.~Chatrchyan {\it et al.}  [CMS Collaboration],
  %``Search for Dark Matter and Large Extra Dimensions in pp Collisions Yielding a Photon and Missing Transverse Energy,''
  Phys.\ Rev.\ Lett.\  {\bf 108}, 261803 (2012)
  [arXiv:1204.0821 [hep-ex]].
  %%CITATION = ARXIV:1204.0821;%%

\bibitem{ATLAS-monoa}
G.~Aad {\it et al.}  [ATLAS Collaboration],
  %``Search for dark matter candidates and large extra dimensions in events with a photon and missing transverse momentum in $pp$ collision data at $\sqrt{s}=7$ TeV with the ATLAS detector,''
  arXiv:1209.4625 [hep-ex].
  %%CITATION = ARXIV:1209.4625;%%
  
  
\bibitem{monogsearches}  
%\cite{Frandsen:2012rk}
%\bibitem{Frandsen:2012rk}
  M.~T.~Frandsen, F.~Kahlhoefer, A.~Preston, S.~Sarkar and K.~Schmidt-Hoberg,
  %``LHC and Tevatron Bounds on the Dark Matter Direct Detection Cross-Section for Vector Mediators,''
  JHEP {\bf 1207} (2012) 123
  [arXiv:1204.3839 [hep-ph]];
  %%CITATION = ARXIV:1204.3839;%%
%\cite{Dreiner:2012xm}
%\bibitem{Dreiner:2012xm}
  H.~Dreiner, M.~Huck, M.~Kramer, D.~Schmeier and J.~Tattersall,
  %``Illuminating Dark Matter at the ILC,''
  arXiv:1211.2254 [hep-ph].
  %%CITATION = ARXIV:1211.2254;%%
  
  
 \bibitem{kimaxion} 
%axion models
%\bibitem{kimaxion}
%\cite{Kim:1979if}
%\bibitem{Kim:1979if}
  J.~E.~Kim,
  %``Weak Interaction Singlet and Strong CP Invariance,''
  Phys.\ Rev.\ Lett.\  {\bf 43} (1979) 103;
  %%CITATION = PRLTA,43,103;%%
%\cite{Shifman:1979if}
%\bibitem{Shifman:1979if}
  M.~A.~Shifman, A.~I.~Vainshtein and V.~I.~Zakharov,
  %``Can Confinement Ensure Natural CP Invariance of Strong Interactions?,''
  Nucl.\ Phys.\  B {\bf 166} (1980) 493.
  %%CITATION = NUPHA,B166,493;%%
 
 
 \bibitem{dineaxion} 
 %\cite{Dine:1981rt}
%\bibitem{Dine:1981rt}
  M.~Dine, W.~Fischler and M.~Srednicki,
  %``A Simple Solution to the Strong CP Problem with a Harmless Axion,''
  Phys.\ Lett.\  B {\bf 104} (1981) 199;
  %%CITATION = PHLTA,B104,199;%%
%\cite{Zhitnitsky:1980tq}
%\bibitem{Zhitnitsky:1980tq}
  A.~R.~Zhitnitsky,
  %``On Possible Suppression of the Axion Hadron Interactions. (In Russian),''
  Sov.\ J.\ Nucl.\ Phys.\  {\bf 31} (1980) 260
  [Yad.\ Fiz.\  {\bf 31} (1980) 497].
  %%CITATION = YAFIA,31,497;%%

  
\bibitem{ROOT}
See webpage: http://root.cern.ch/.

 \bibitem{Feynrules}
 N.~D.~Christensen and C.~Duhr,
 {\it FeynRules - Feynman rules made easy,}
  Comput.\ Phys.\ Commun.\  {\bf 180} (2009) 1614
  [arXiv:0806.4194 [hep-ph]].
  %%CITATION = ARXIV:0806.4194;%%
  

\bibitem{MG5} 
  J.~Alwall {\it et al.}
{\it MadGraph 5 : Going Beyond,}
  JHEP {\bf 1106}, 128 (2011)
  [arXiv:1106.0522 [hep-ph]].
  %%CITATION = ARXIV:1106.0522;%%
  
  \bibitem{UFO}
 C.~Degrande, C.~Duhr, B.~Fuks, D.~Grellscheid, O.~Mattelaer and T.~Reiter,
{\it UFO - The Universal FeynRules Output,}
  Comput.\ Phys.\ Commun.\  {\bf 183} (2012) 1201
  [arXiv:1108.2040 [hep-ph]].
  %%CITATION = ARXIV:1108.2040;%%
  
\bibitem{PYTHIA} 
  T.~Sjostrand, S.~Mrenna and P.~Z.~Skands,
  {\it PYTHIA 6.4 Physics and Manual,}
  JHEP {\bf 0605}, 026 (2006)
  [hep-ph/0603175].
  %%CITATION = HEP-PH/0603175;%%
  
\bibitem{Delphes} 
  S.~Ovyn, X.~Rouby and V.~Lemaitre,
  {\it DELPHES, a framework for fast simulation of a generic collider experiment,}
  arXiv:0903.2225 [hep-ph].
  %%CITATION = ARXIV:0903.2225;%%





\bibitem{Hdiphoton}
%\cite{Joglekar:2012hb}
%\bibitem{Joglekar:2012hb}
  A.~Joglekar, P.~Schwaller and C.~E.~M.~Wagner,
  %``Dark Matter and Enhanced Higgs to Di-photon Rate from Vector-like Leptons,''
  arXiv:1207.4235 [hep-ph];
 %%CITATION = ARXIV:1207.4235;%%
%\cite{Carena:2012xa}
%\bibitem{Carena:2012xa}
  M.~Carena, I.~Low and C.~E.~M.~Wagner,
  %``Implications of a Modified Higgs to Diphoton Decay Width,''
  arXiv:1206.1082 [hep-ph];
  %%CITATION = ARXIV:1206.1082;%%
%\cite{ArkaniHamed:2012kq}
%\bibitem{ArkaniHamed:2012kq}
  N.~Arkani-Hamed, K.~Blum, R.~T.~D'Agnolo and J.~Fan,
  %``2:1 for Naturalness at the LHC?,''
  arXiv:1207.4482 [hep-ph];
  %%CITATION = ARXIV:1207.4482;%%
%\cite{Almeida:2012he}
%\bibitem{Almeida:2012he}
  L.~G.~Almeida, E.~Bertuzzo, P.~A.~N.~Machado and R.~Z.~Funchal,
  %``Does $H \to \gamma \gamma$ Taste like vanilla New Physics?,''
  arXiv:1207.5254 [hep-ph];
  %%CITATION = ARXIV:1207.5254;%%
%\cite{Kearney:2012zi}
%\bibitem{Kearney:2012zi}
  J.~Kearney, A.~Pierce and N.~Weiner,
  %``Vectorlike Fermions and Higgs Couplings,''
  arXiv:1207.7062 [hep-ph];
  %%CITATION = ARXIV:1207.7062;%%
%\cite{Batell:2012zw}
%\bibitem{Batell:2012zw}
  B.~Batell, S.~Jung and H.~M.~Lee,
  %``Singlet Assisted Vacuum Stability and the Higgs to Diphoton Rate,''
  arXiv:1211.2449 [hep-ph].
  %%CITATION = ARXIV:1211.2449;%%





\bibitem{antiproton}
%\cite{Adriani:2010rc}
%\bibitem{Adriani:2010rc}
  O.~Adriani {\it et al.}  [PAMELA Collaboration],
  %``PAMELA results on the cosmic-ray antiproton flux from 60 MeV to 180 GeV in kinetic energy,''
  Phys.\ Rev.\ Lett.\  {\bf 105} (2010) 121101
  [arXiv:1007.0821 [astro-ph.HE]];
  %%CITATION = ARXIV:1007.0821;%%
%\cite{Belanger:2012ta}
%\bibitem{Belanger:2012ta}
  G.~Belanger, C.~Boehm, M.~Cirelli, J.~Da Silva and A.~Pukhov,
  %``PAMELA and FERMI-LAT limits on the neutralino-chargino mass degeneracy,''
  JCAP {\bf 1211} (2012) 028
  [arXiv:1208.5009 [hep-ph]].
  %%CITATION = ARXIV:1208.5009;%%



\bibitem{continuum}
%\cite{Chu:2012qy}
%\bibitem{Chu:2012qy}
  X.~Chu, T.~Hambye, T.~Scarna and M.~H.~G.~Tytgat,
  %``What if Dark Matter Gamma-Ray Lines come with Gluon Lines?,''
  arXiv:1206.2279 [hep-ph];
  %%CITATION = ARXIV:1206.2279;%%
%\cite{Buchmuller:2012rc}
%\bibitem{Buchmuller:2012rc}
  W.~Buchmuller and M.~Garny,
  %``Decaying vs Annihilating Dark Matter in Light of a Tentative Gamma-Ray Line,''
  JCAP {\bf 1208} (2012) 035
  [arXiv:1206.7056 [hep-ph]];
  %%CITATION = ARXIV:1206.7056;%%
%\cite{Cohen:2012me}
%\bibitem{Cohen:2012me}
  T.~Cohen, M.~Lisanti, T.~R.~Slatyer and J.~G.~Wacker,
  %``Illuminating the 130 GeV Gamma Line with Continuum Photons,''
  arXiv:1207.0800 [hep-ph];
  %%CITATION = ARXIV:1207.0800;%%
%\cite{Cholis:2012fb}
%\bibitem{Cholis:2012fb}
  I.~Cholis, M.~Tavakoli and P.~Ullio,
  %``Searching for the continuum spectrum photons correlated to the 130 GeV gamma-ray line,''
  arXiv:1207.1468 [hep-ph];
  %%CITATION = ARXIV:1207.1468;%%
  %\cite{Huang:2012yf}
%\bibitem{Huang:2012yf}
  X.~-Y.~Huang, Q.~Yuan, P.~-F.~Yin, X.~-J.~Bi and X.~-L.~Chen,
  %``Constraints on the dark matter annihilation scenario of Fermi 130 GeV $\gamma$-ray line emission by continuous gamma-rays, Milky Way halo, galaxy clusters and dwarf galaxies observations,''
  arXiv:1208.0267 [astro-ph.HE].
  %%CITATION = ARXIV:1208.0267;%%





\bibitem{monoa-vero}
G.~Belanger, M.~Heikinheimo and V.~Sanz,
  %``Model-Independent Bounds on Squarks from Monophoton Searches,''
  JHEP {\bf 1208}, 151 (2012)
  [arXiv:1205.1463 [hep-ph]].
  %%CITATION = ARXIV:1205.1463;%%




\bibitem{MVX}
 J.~Ellis, D.~S.~Hwang, V.~Sanz and T.~You,
  %``A Fast Track towards the `Higgs' Spin and Parity,''
  arXiv:1208.6002 [hep-ph].
  %%CITATION = ARXIV:1208.6002;%%




\bibitem{VBF}
G.~L.~Kane, W.~W.~Repko and W.~B.~Rolnick,
  %``The Effective W+-, Z0 Approximation for High-Energy Collisions,''
  Phys.\ Lett.\ B {\bf 148}, 367 (1984).
  %%CITATION = PHLTA,B148,367;%%
  J.~Bagger, V.~D.~Barger, K.~-m.~Cheung, J.~F.~Gunion, T.~Han, G.~A.~Ladinsky, R.~Rosenfeld and C.~P.~Yuan,
  %``The Strongly interacting W W system: Gold plated modes,''
  Phys.\ Rev.\ D {\bf 49} (1994) 1246
  [hep-ph/9306256].
  %%CITATION = HEP-PH/9306256;%%
   J.~Bagger, V.~D.~Barger, K.~-m.~Cheung, J.~F.~Gunion, T.~Han, G.~A.~Ladinsky, R.~Rosenfeld and C.~-P.~Yuan,
  %``CERN LHC analysis of the strongly interacting W W system: Gold plated modes,''
  Phys.\ Rev.\ D {\bf 52}, 3878 (1995)
  [hep-ph/9504426].
  %%CITATION = HEP-PH/9504426;%%


\bibitem{LHCmonoZ}
%\cite{:2012kg}
%\bibitem{:2012kg}
  G.~Aad {\it et al.}  [ATLAS Collaboration],
  %``Measurement of ZZ production in pp collisions at sqrt(s)=7 TeV and limits on anomalous ZZZ and ZZgamma couplings with the ATLAS detector,''
  arXiv:1211.6096 [hep-ex].
  %%CITATION = ARXIV:1211.6096;%%

\bibitem{mono-Z}
 Y.~Bai and T.~M.~P.~Tait,
  %``Searches with Mono-Leptons,''
  arXiv:1208.4361 [hep-ph].
  %%CITATION = ARXIV:1208.4361;%%
N.~F.~Bell, J.~B.~Dent, A.~J.~Galea, T.~D.~Jacques, L.~M.~Krauss and T.~J.~Weiler,
  %``Searching for Dark Matter at the LHC with a Mono-Z,''
  arXiv:1209.0231 [hep-ph];
  %%CITATION = ARXIV:1209.0231;%%
%\cite{Carpenter:2012rg}
%\bibitem{Carpenter:2012rg}
  L.~M.~Carpenter, A.~Nelson, C.~Shimmin, T.~M.~P.~Tait and D.~Whiteson,
  %``Collider searches for dark matter in events with a Z boson and missing energy,''
  arXiv:1212.3352 [hep-ex].
  %%CITATION = ARXIV:1212.3352;%%


\bibitem{LEP3gamma}
%\cite{Abbiendi:2002je}
%\bibitem{Abbiendi:2002je}
  G.~Abbiendi {\it et al.}  [OPAL Collaboration],
  %``Multiphoton production in e+ e- collisions at s**(1/2) = 181-GeV to 209-GeV,''
  Eur.\ Phys.\ J.\ C {\bf 26} (2003) 331
  [hep-ex/0210016].
  %%CITATION = HEP-EX/0210016;%%



\bibitem{HB}
G. Raffelt, Stars as laboratories for fundamental physics: The astrophysics of neutrinos, axions, and other weakly interacting particles. University of Chicago Press, Chicago IL, US, 1996.





\end{thebibliography}
\end{document}